\documentclass[acus]{JAC2003}
\usepackage{graphicx}

\renewcommand{\bar}[1]{\overline{#1}}
\newcommand{\VEV}[1]{\left\langle{#1}\right\rangle}
\newcommand{\eg}{{\em e.g.}}
\newcommand{\ket}[1]{\vert\,{#1}\rangle}
\newcommand{\M}{\mathcal{M}}

\newcommand{\MSbar} {\hbox{$\overline{\hbox{\tiny MS}}$}}
\newcommand{\barMS} {\overline{\rm MS}}

\begin{document}
\title{QCD PHYSICS OPPORTUNITIES IN LOW-ENERGY ELECTRON-POSITRON
ANNIHILATION\thanks{This work was supported by the Department of Energy
contract DE-AC03-76SF00515}}

\author{Stanley J. Brodsky\thanks{sjbth@slac.stanford.edu}, SLAC,
Stanford, CA 94305, USA}

\maketitle

\begin{abstract}

I survey a number of interesting tests of quantum chromodynamics
at the amplitude level which can be carried out in
electron-positron annihilation and in photon-photon collisions at
low energy.  Some of the tests require $e^+ e^-$ center of mass
energy as small  as $\sqrt s = 2 $ GeV.  Other tests which involve
a spectrum of energies can be carried out advantageously at high
energy facilities using the radiative return method. These include
measurements of fundamental processes such as timelike form
factors and transition amplitudes, timelike Compton scattering,
timelike  photon to meson transition amplitudes, and two-photon
exclusive processes.  Many of these reactions test basic
principles of QCD such as hadronization at the amplitude level,
factorization, and hadron helicity conservation, tools also used
in the analysis of exclusive $B$ and $D$ decays. Measurements of
the final-state  polarization in hadron pair production determine
the relative phase of the timelike form factors and thus strongly
discriminate between analytic forms of models which fit the  form
factors in the spacelike region.  The role of two-photon exchange
amplitudes can be tested using the charge asymmetry of the $e^+
e^- \to B \bar B$ processes.  These tests can help resolve the
discrepancy between the Jefferson laboratory measurements of the
ratio of $G_E$ and $G_M$ proton form factors using the
polarization transfer method versus measurements using the
traditional Rosenbluth method. Precision measurements of the
electron-positron annihilation cross section can test the
generalized Crewther relation and determine whether the effective
couplings defined from physical measurements show infrared
fixed-point and near conformal behavior. I also discuss a number
of tests of novel QCD phenomena accessible in $e^+ e^-$
annihilation, including near-threshold reactions, the production
of baryonium, gluonium states, and pentaquarks.

\end{abstract}

\section{INTRODUCTION}

Quantum Chromodynamics has been very well tested at high energies,
particularly in inclusive reactions involving large momentum
transfers much higher than the QCD scale $\Lambda_{QCD}.$ Tests of
QCD  at low energies  are much more challenging, since they
require an understanding of nonperturbative elements of the
theory, including the behavior of the QCD coupling at low momentum
transfers, the fundamental features of hadron wavefunctions, and
the fundamental color coherence of QCD interactions.

In this talk I will survey a number of tests of QCD which test
fundamental issues of hadron physics at the amplitude level.
These include measurements of fundamental processes such as
timelike form factors and transition amplitudes, timelike Compton
scattering, timelike  photon to meson transition amplitudes, and
two-photon exclusive processes.  Many of these reactions test
basic principles of QCD such as factorization and hadron helicity
conservation, tools also used in the analysis of exclusive $B$ and
$D$ decays. Electron-positron annihilation can determine whether
the effective couplings defined from physical measurements show
infrared fixed-point and near-conformal behavior.  I also discuss
a number of tests of novel QCD phenomena accessible in $e^+ e^-$
annihilation, including near-threshold reactions, the production
of baryonium, gluonium states, and pentaquarks.

There has recently been a number of experimental surprises in QCD
spectroscopy and heavy quark production which show the importance
of detailed measurements in electron-positron collisions. These
include:

1.  The cross section measured at Belle~\cite{Abe:2002rb} for
double-charmonium production $e^+ e^- \to J/\psi \eta_c$ and
$J/\psi D X$ is an order of magnitude larger than
predicted~\cite{Ioffe:2003ru}, It is important to see whether this
anomaly also holds for analogous channels involving strangeness:
$e^+ e^- \to \phi \eta$ and  $\phi K X.$

\begin{figure}[htb]
\begin{center}
\includegraphics*[width=75mm]{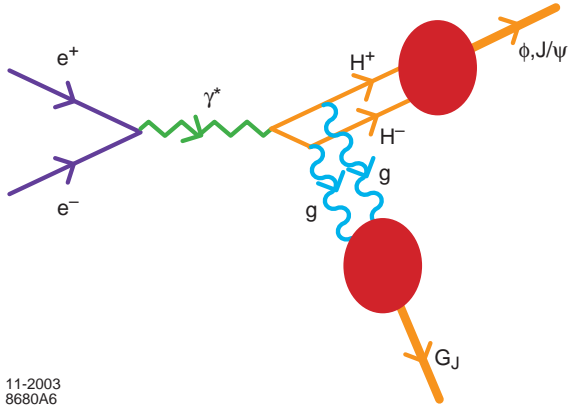}
\end{center}
\caption[*]{Illustration of QCD mechanism for the exclusive production of
quarkonium and gluonium in electron-positron collisions.}
\label{sar8680A6}
\end{figure}

2.  The evidence for the predicted gluonic bound states $gg$,
$ggg$, $q \bar q g$ spectroscopy of QCD is still not conclusive.
The subprocesses $e^+ e^- \to c \bar c g g $ and $e^+ e^- \to c
\bar c c \bar c$ have comparable rates, suggesting a large role
for the production of associated charmonium plus gluonic states.
See Fig. \ref{sar8680A6}.  Fred Goldhaber, Jungil Lee and I have
recently calculated the cross section for $e^+e^-\to H
\mathcal{G}_{J=0,2}$ using perturbative QCD
factorization~\cite{Brodsky:2003hv}. We find that  $\gamma^* \to
J/\psi \mathcal{G}_0$ production dominates over that of $J/\psi
\mathcal{G}_2$, and show how the angular distribution of the final
state can be used to determine the angular momentum $J$ and
projection $J_z$ of the glueball; only $J_z = \pm 2$ tensor states
are produced by the perturbative QCD mechanism at leading twist.
The rate for $e^+e^-\to J/\psi \mathcal{G}_0$ production could be
comparable to the corresponding nonrelativistic QCD (NRQCD)
prediction for $e^+e^-\to J/\psi\eta_c$ without exceeding the
known bound from radiative $\Upsilon$ decay.  Another interesting
glueball search process is the missing mass spectrum in $e^+ e^-
\to \phi X.$

3.  New signals for baryonium resonances just below threshold in $p \bar p \to
e^+ e^- $ (odd charge conjugation) and $J/\psi \to (p \bar p) \gamma$ (even
charge conjugation) have been reported~\cite{Bai:2003sw,Abe:2002ds,Abe:2002tw},
but not in $J/\psi \to (p \bar p) \pi^0$ (odd charge conjugation).  A  strong
threshold enhancement is also observed in $p \bar p \to e^+
e^-$~\cite{Andreotti:bt}.  These near-threshold states may reflect the binding of
$qqq$ or $qq$ systems, the QCD van der Waals
interaction~\cite{Brodsky:1989jd,Luke:1992tm}, quark interchange covalent bonds,
or attractive meson exchange interactions analogous to the nuclear
potential~\cite{Richard:1999qh,Datta:2003iy}.  It is important to test this
phenomenon not only in baryon production near threshold $e^+ e^- \to B \bar B$
and $\gamma \gamma \to B \bar B,$ but also for any hadron-pair threshold. It is
also interesting to study the formation of Coulomb-bound atomic states such as
$\mu^+ \mu^-$ and $\tau^+ \tau^-.$

\begin{figure}[htb]
\begin{center}
\includegraphics*[width=75mm]{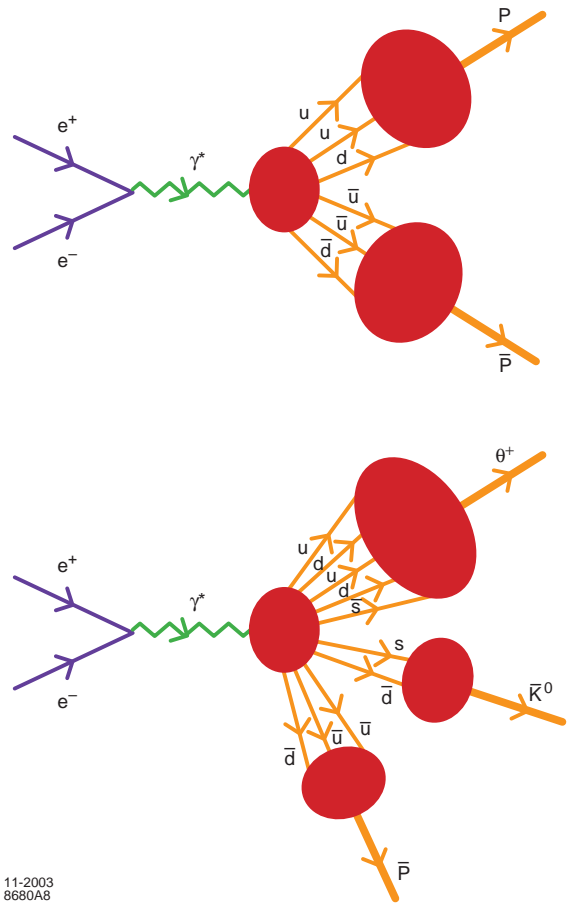}
\end{center}
\caption[*]{Mechanisms for producing baryon pairs and pentaquarks}
\label{sar8680A08}
\end{figure}

4.  The recent
discovery~\cite{Nakano:2003qx,Barmin:2003vv,Stepanyan:2003qr,Barth:2003es}
of a pentaquark state $\Theta^+(udd u \bar s)$ indicates that the
spectroscopy of QCD is much richer than previously
thought~\cite{Diakonov:1997mm,Close:2003tv,Karliner:2003sy,Roy:2003hk,%
Carlson:2003pn} These states could be produced and analyzed in
exclusive reactions such as $e^+ e^- \to \Theta^+(uudd\bar s)
\Theta^-(\overline {u  u  d d} s)$ and  $e^+ e^- \to
\Theta^+(uudd\bar s) K^-(\bar u s) \bar n(\bar u \bar d \bar d) $
or $\Theta^+ \bar K^0 \bar p$.  See Fig. \ref{sar8680A08}. These
types or reactions can give decisive information on the quantum
numbers of the new states. The form factor for timelike pentaquark
pair production is predicted to falloff as $s^{-4}$ according to
dimensional counting rules.

5.  The transition form factor for $e p \to e \Delta^+ $ falls anomalously fast
at high spacelike $q^2$ compared to other  $e p \to e N^* $ form
factors~\cite{Stoler:1993yk}.  Is this due to a special characteristic of
$\Delta$ substructure?  For example, if the $\Delta^+$ is dominantly a pentaquark
$uuuu\bar d$ state, then the timelike $e^+ e^- \to \bar p \Delta^+$ transition
form factor will fall as ${s}^{-3}$ compared to the canonical ${s}^{-2}$ QCD
fall-off for timelike baryon pair form factors.

6.  The phenomenology of $J/\psi$ decays is still
puzzling~\cite{Chernyak:1999cj}. For example, the decay $J/\psi
\to \rho \pi$ is the largest two-body decay channel for the
$J/\psi$ even though the decay of the $J/\psi$ with $J^z = \pm 1$
to pseudoscalar vector channels is forbidden by hadron helicity
conservation~\cite{Brodsky:1981kj}, which follows from QCD
factorization and the chirality conservation of QCD interactions.
In contrast, the $\psi^\prime$ has not been observed to decay to
$\rho \pi.$  As discussed below, an intriguing solution of this
puzzle is the presence of intrinsic heavy quark Fock states in the
light hadrons~\cite{Brodsky:1997fj}.  It is important to verify
whether hadron-helicity conservation is observed in the continuum
production of meson pairs; \eg, $e^+ e^- \to \rho \pi$ should be
strongly suppressed.

7.  The form factors of hadrons as measured in both the spacelike and
timelike domains provide fundamental information on the structure
and internal dynamics of hadrons.  Recent
measurements~\cite{perdrisat} of the electron-to-proton
polarization transfer in $\overrightarrow e^- \, p \to e^- \,
\overrightarrow p$ scattering at Jefferson Laboratory show that
the ratio of spacelike Sachs form factors~\cite{walecka} $G^p_E(q^2)/
G^p_M(q^2)$ is monotonically decreasing with increasing
$Q^2=-q^2,$ in strong contradiction with the $G_E/G_M$ scaling
determined by the traditional Rosenbluth separation method.  The
Rosenbluth method may in fact not be reliable, perhaps because of
its sensitivity to uncertain radiative corrections, including
two-photon exchange amplitudes~\cite{guichon}.  The polarization
transfer method~\cite{perdrisat,acg81} is relatively insensitive
to such corrections.

The same data which indicate that $G_E$ for protons falls faster
than $G_M$ at large spacelike $Q^2$ require in turn that $F_2/F_1$
falls more slowly than $1/Q^2$.  The conventional expectation from
dimensional counting rules~\cite{Brodsky:1974vy} and perturbative
QCD~\cite{Lepage:1979za} is that the Dirac form factor $F_1$
should fall with a nominal power $1/Q^4$, and the ratio of the
Pauli and Dirac form factors, $F_2/F_1$, should fall like $1/Q^2$,
at high momentum transfers.  The Dirac form factor agrees with this
expectation in the range $Q^2$ from a few GeV$^2$ to the data
limit of 31 GeV$^2$.  However, the Pauli/Dirac ratio is not
observed to fall with the nominal expected power, and the
experimenters themselves have noted that the data is well fit by
$F_2/F_1 \propto 1/Q$ in the momentum transfer range 2 to 5.6
GeV$^2$.

The new Jefferson Laboratory results make it critical to carefully
identify and separate the timelike $G_E$ and $G_M$ form factors by
measuring the center-of-mass angular distribution and by measuring
the polarization of the proton in baryon pair $e^+ e^- \to B \bar
B$ reactions.  Polarization measurements can determine phase
structure of the timelike form factors and thus provide a
remarkable window into QCD at the amplitude
level~\cite{Brodsky:2003gs}.  The role of two-photon exchange
amplitudes can be tested using the charge asymmetry of the $e^+
e^- \to B \bar B$ processes.  The advent of high luminosity $e^+
e^-$ colliders at Frascati and elsewhere provide the opportunity
to make such measurements, both directly and via radiative return.

The advent of electron-positron colliders of high luminosity thus
can open up a new range of sensitive tests of QCD.  The following
sections give an introduction to the theory of a number of $e^+
e^-$ collider topics including hard exclusive processes such as
timelike Compton scattering, timelike photon-to-pion transition
amplitudes, two-photon exclusive processes and single-spin
polarization asymmetries.  Many of these topics test the main
tools used in the analysis of exclusive $B$ and $D$ decays and
thus are highly relevant to progress in that field.

\begin{figure}[htb]
\begin{center}
\includegraphics*[width=75mm]{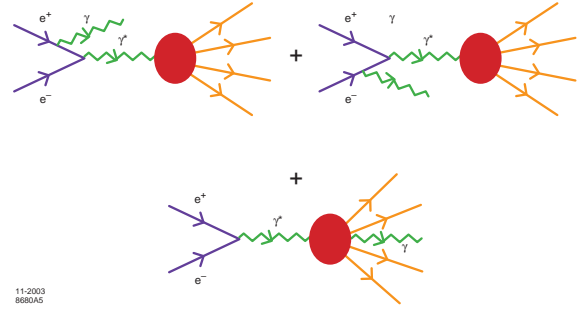}
\end{center}
\caption[*]{Illustration of initial state radiation in electron-positron
collisions.  The contribution where the photon is emitted from the hadron
currents causes charge and spin asymmetries.}
\label{sar8680A5}
\end{figure}

Some of the tests discussed here require $e^+ e^-$ center of mass
energy as small as $\sqrt s = 2 $ GeV.  Other tests which involve
a spectrum of energies can be carried out advantageously at high
energy facilities using initial-state
radiation~\cite{Aloisio:2001xq,Solodov:2002xu,Aubert:2003sv}. See
Fig. ~\ref{sar8680A5}. Recent results from KLOE and BaBar are
given in~\cite{Aloisio:2001xq,Solodov:2002xu,Aubert:2003sv}. The
radiation of a hard photon from the initial-state electron or the
positron allows one to measure the annihilation cross section at a
lower energy $s(1-x).$ The basic formula is:
\begin{equation}
{d \sigma(s,x)\over dx } = W(s,x) \sigma[s(1-x)]\end{equation}
where in Born approximation \begin{equation} W(s,x) =
{2\alpha\over\pi x}(2\ln{\sqrt{s\over m_e}} -1)(1-x+{x^2\over
2})
\end{equation}
Although the effective luminosity using ISR is reduced by the
probability for radiation, this is compensated by the fact that
one measures the entire spectrum at one setting of $s$.  The hard
photons are primarily radiated along the initial lepton direction.
Half of the radiative cross section occurs at $\theta \le
\sqrt{m_e\over E_e}.$  If the photon is radiated at large angles,
then one can test for single spin asymmetries relative to the
normal $\vec n = \vec p_{e^+} \times \vec p_{e^-}$ to the
annihilation plane.

In radiative inclusive reactions, $e^+ e^- \to q \bar q \gamma$,
the interference of initial and final state radiation produces jet
charge asymmetries which measure the cube of the quark charge
$e^3_q.$~\cite{Brodsky:1976fp,Brodsky:1972yx} The hadron charge
asymmetry in semi-inclusive reactions $e^+ e^- \to H^\pm \gamma X$
determines interesting odd charge-conjugation fragmentation
functions~\cite{Brodsky:1976fp}. In the case of exclusive
reactions, one can use the hadron charge asymmetry to measure the
interference of the timelike Compton amplitude $e^+ e^- \to
\gamma^* \to H^+ H^-\gamma $ with the timelike form factor
appearing in the initial state radiation amplitude $e^+ e^- \to
\gamma^* \gamma \to H^+ H^- \gamma.$ This remarkable physics
potential is discussed in more detail below. The corresponding
formulae for lepton charge asymmetries in $e^+ e^- \to \ell \bar
\ell \gamma$ are given in Ref.~\cite{Brodsky:1976fp}

\section{EXCLUSIVE PROCESSES IN QCD}

Hard hadronic exclusive processes such as timelike annihilation
$e^+ e^- \to H \bar H$ and $\gamma \gamma \to H \bar H$ are at the
forefront of low energy QCD studies, particularly because of their
role in the interpretation of exclusive hadronic $B$ decays,
processes which are essential for determining the CKM phases and
the physics of $CP$ violation.  Perturbative QCD and its
factorization properties at high momentum transfer provide an
essential guide to the phenomenology of exclusive amplitudes at
large momentum transfer---the leading power fall-off of form
factors and fixed-angle cross sections, the dominant helicity
structures, and their color transparency properties.  The
perturbative QCD predictions for two photon reactions can be
compared with a phenomenological successful model based on the
handbag approximation~\cite{Kroll:2003cd}.

A primary issue is the nature and shapes of hadron light-front
wavefunctions, the amplitudes which interpolate between hadrons
and their quark and gluon degrees of freedom.  This is particular
important for $B$ physics since the calculation of exclusive
hadronic $B$ are computed from the convolution of hadron
wavefunctions and distribution amplitudes.  For example, the decay
amplitude for $B \to \ell \bar \nu \pi$ is exactly given by the
overlap of $B$ and $\pi$  light-front wavefunctions.  Furthermore
the phase structure of hadronic amplitudes and the effects of
color transparency are directly relevant to the analysis of phase
structure of $B$ decays.

There has been considerable progress analyzing exclusive and
diffractive reactions at large momentum transfer from first
principles in QCD.  Rigorous statements can be made on the basis
of asymptotic freedom and factorization theorems which separate
the underlying hard quark and gluon subprocess amplitude from the
nonperturbative physics of the hadronic wavefunctions.  The
leading-power contribution to exclusive hadronic amplitudes such
as quarkonium decay, heavy hadron decay,  and scattering
amplitudes involving large momentum transfer can usually be
factorized as a  convolution of distribution amplitudes
$\phi_H(x_i,\Lambda)$ and hard-scattering quark/gluon scattering
amplitudes $T_H$ integrated over the light-cone momentum fractions
of the valence quarks~\cite{Lepage:1980fj}:
\begin{eqnarray}
\M_{\rm Hadron} &=&\int
 \prod \phi_H^{(\Lambda)} (x_i,\lambda_i)\, T_H^{(\Lambda)} dx_i\ .
\label{eq:e}
\end{eqnarray}
Here $T_H^{(\Lambda)}$ is the underlying quark-gluon
scattering amplitude subprocess in which each incident and final hadron is
replaced by valence quarks with collinear momenta $k^+_i =x_i
p^+_H$, $\vec k_{\perp i} = x_i \vec p_{\perp H }.$ The invariant
mass of all intermediate states in $T_H$ is evaluated above the
separation scale $\M^2_n > \Lambda^2$.  The essential part of the
hadronic wavefunction is the distribution amplitude
\cite{Lepage:1980fj}, defined as the integral over transverse
momenta of the valence (lowest particle number) Fock wavefunction;
\eg\ for the pion
\begin{equation}
\phi_\pi (x_i,Q) \equiv \int d^2k_\perp\, \psi^{(Q)}_{q\bar q/\pi}
(x_i, \vec k_{\perp i},\lambda) \label{eq:f}
\end{equation}
where the separation scale $\Lambda$ can be taken to be order of
the characteristic momentum transfer $Q$ in the process.  It
should be emphasized that the hard scattering amplitude $T_H$ is
evaluated in the QCD perturbative domain where the propagator
virtualities are above the separation scale.  The leading power
fall-off of the hard-scattering amplitude as given by dimensional
counting rules follows from the nominal scaling of the
hard-scattering amplitude: $T_H \sim 1/Q^{n-4}$, where $n$ is the
total number of fields (quarks, leptons, or gauge fields)
participating in the hard
scattering~\cite{Brodsky:1974vy,Matveev:1973ra}.  Thus the
reaction is dominated by subprocesses and Fock states involving
the minimum number of interacting fields.  In the case of $2 \to
2$ scattering amplitudes, this implies $ {d\sigma\over dt}(A B \to
C D) ={F_{A B \to C D}(t/s)/ s^{n-2}}.$ In the case of form
factors, the dominant helicity conserving amplitude obeys $F(t)
\sim (1/t)^{n_H-1}$ where $n_H$ is the minimum number of fields in
the hadron $H$.  The full predictions from PQCD modify the nominal
scaling by logarithms from the running coupling and the evolution
of the distribution amplitudes.  In some cases, such as large
angle $pp \to p p $ scattering, there can be ``pinch"
contributions~\cite{Landshoff:ew} when the scattering can occur
from a sequence of independent near-on shell quark-quark
scattering amplitudes at the same CM angle.  After inclusion of
Sudakov suppression form factors, these contributions also have a
scaling behavior close to that predicted by constituent counting.

As shown by Maldacena~\cite{Maldacena:1997re}, there is a
remarkable correspondence between large $N_C$ supergravity theory
in a higher dimensional  anti-de Sitter space  and supersymmetric
QCD in 4-dimensional space-time.  String/gauge duality provides a
framework for predicting QCD phenomena based on the conformal
properties of the ADS/CFT correspondence.  In a remarkable recent
development, Polchinski and Strassler~\cite{Polchinski:2001tt}
have derived the  dimensional counting rules using string duality,
mapping features of gravitational theories in higher dimensions
$(AdS_5)$ to physical QCD in ordinary 3+1 space-time.  The
power-law fall-off of hard exclusive hadron-hadron scattering
amplitudes at large momentum transfer can be derived without the
use of perturbation theory by using the scaling properties of the
hadronic interpolating fields in the large-$r$ region of  AdS
space.

The distribution amplitudes which control leading-twist exclusive
amplitudes at high momentum transfer can be related to the
gauge-invariant Bethe-Salpeter wavefunction at equal light-cone
time $\tau = x^+$.  The logarithmic evolution of the hadron
distribution amplitudes $\phi_H (x_i,Q)$ with respect to the
resolution scale $Q$ can be derived from the
perturbatively-computable tail of the valence light-cone
wavefunction in the high transverse momentum regime.  The DGLAP
evolution of quark and gluon distributions can also be derived in
an analogous way by computing the variation of the Fock expansion
with respect to the separation scale.  Other key features of the
perturbative QCD analyses are: (a) evolution equations for
distribution amplitudes which incorporate the operator product
expansion, renormalization group invariance, and conformal
symmetry~\cite{Lepage:1980fj,Brodsky:1980ny,Muller:1994cn,Ball:1998ff,%
Braun:1999te}; (b) hadron helicity conservation which follows from
the underlying chiral structure of QCD~\cite{Brodsky:1981kj}; (c)
color transparency, which eliminates corrections to hard exclusive
amplitudes from initial and final state interactions at leading
power and reflects the underlying gauge theoretic basis for the
strong interactions~\cite{Brodsky:1988xz} and (d) hidden color
degrees of freedom in nuclear wavefunctions, which reflect the
color structure of hadron and nuclear
wavefunctions~\cite{Brodsky:1983vf}.  There have also been recent
advances eliminating renormalization scale ambiguities in
hard-scattering amplitudes via commensurate scale
relations~\cite{Brodsky:1994eh} which connect the couplings
entering exclusive amplitudes to the $\alpha_V$ coupling which
controls the QCD heavy quark potential.

\section{EXCLUSIVE TIMELIKE REACTIONS AND HADRON HELICITY CONSERVATION}

Measurements of exclusive hadronic amplitudes in the timelike
domain can test many basic principles of QCD, including
factorization principles, dimensional counting rules, hadron
helicity conservation, color transparency and the possible role of
higher Fock states such as intrinsic charm.  Dimensional counting
rules test the near conformal nature of QCD at moderate to high
momentum transfers.  The essential prediction for the production
cross section of $N$ hadrons each emitted at distinct fixed CM
angles is the leading power-law prediction
\begin{equation}
{dR_{e^+ e^- \to H_1 H_2  ... H_N}\over d\Omega_1 d\Omega_2  ... d
\Omega_{N-1}}(s) \propto \big[{\alpha_s \Lambda_{QCD}^2\over
s}\big]^{n_{tot}-2},
\end{equation}
where $n_{tot}$ is the total number of quark and gluon
constituents in the final state hadrons.  The prediction is
modified by possible anomalous dimensions and the running of the
QCD coupling.  However, there is now substantial theoretical and
empirical evidence that the QCD coupling has an effective IR fixed
point and can be treated as a constant over a large range of
momentum transfers.  In the case of two-body final states, this
scaling predicts $s F_H(s) \to {\rm const}$ for meson pairs and
$s^2 F_H(s) \to {\rm const}$ for baryon pairs, $s^5 F(s) \to {\rm
const} $ for deuteron pairs, and $s^4 F(s) \to {\rm const} $ for
pentaquark pairs.  As discussed in the introduction, the anomalous
fall-off of the proton to $\Delta$ transition form factor may
indicate a dominance of higher Fock states in the $\Delta.$ This
can be tested by measuring the power-law  fall-off of $e^+ e^- \to
\bar p \Delta.$

Hadron helicity conservation (HHC) is a QCD selection rule
concerning the behavior of helicity amplitudes at high momentum
transfer, such as fixed CM scattering.  Since the convolution of
$T_H$ with the light-cone wavefunctions projects out states with
$L_z=0$, the leading hadron amplitudes conserve hadron
helicity~\cite{Brodsky:1981kj,Chernyak:1999cj}.  Thus the dominant
amplitudes are those in which the sum of hadron helicities in the
initial state equals the sum of hadron helicities in the final
state; other helicity amplitudes are relatively suppressed by an
inverse power in the momentum transfer.

The study of time-like hadronic form factors using $e^+ e^-$
colliding beams can provide very sensitive tests of hadron helicity
conservation, since the virtual photon in $e^+ e^- \to \gamma^* \to h_A
\bar h_B$ always has spin $\pm 1$ along the beam axis at high energies.
Angular momentum conservation implies that the virtual photon can ``decay"
with one of only two possible angular distributions in the center
of momentum frame: $(1+ \cos^2\theta)$ for $\vert\lambda_A -
\lambda_B \vert = 1$ and $\sin^2 \theta$ for $\vert \lambda_A -
\lambda_B \vert = 0$ where $\lambda_A$ and $\lambda_B$ are the
helicities of the outgoing hadrons.  Hadronic helicity
conservation, as required by QCD, greatly restricts the
possibilities.  It implies that $\lambda_A + \lambda_B = 0$.
Consequently, angular momentum conservation requires $\vert
\lambda_A\vert = \vert \lambda_B \vert = l/2$ for baryons, and
$\vert \lambda_A\vert = \vert \lambda_B \vert = 0$ for mesons;
thus the angular distributions for any sets of hadron pairs are
now completely determined at leading twist: $ {d \sigma \over d
\cos\theta}(e^+ e^- = B \bar B) \propto 1 + \cos^2 \theta $ and $
{d \sigma \over d \cos \theta} (e^+ e^- = M \bar M) \propto \sin^2
\theta  .$ Verifying these angular distributions for vector mesons
and other higher spin mesons and baryons would verify the vector
nature of the gluon in QCD and the validity of PQCD applications
to exclusive reactions.  In the case of vector pseudoscalar
channels, parity conservation requires that the vector meson has
$J_z = \pm 1.$ Thus the vector-pseudoscalar meson pairs must be
suppressed in the leading twist limit; {\em e.g.}
\begin{equation}
{\sigma_{e^+ e^-
\to \rho \pi}(s)\over \sigma_{e^+ e^- \to \pi^+ \pi^-}(s)}
\propto{\Lambda^2_{QCD}\over s} \ .
\end{equation}
Surprisingly, this critical PQCD prediction  has not been tested.
If it fails, the perturbative QCD approach to hard exclusive  hadron
processes including the QCD factorization predictions for
exclusive $B$ decays would be in question.

In the case of electron-proton scattering, hadron helicity
conservation states that the proton helicity-conserving form
factor ( which is proportional to $G_M$) dominates over the proton
helicity-flip amplitude  (proportional to $G_E/\sqrt \tau $) at
large momentum transfer.  Here $\tau = Q^2/4M^2, Q^2 = -q^2.$ Thus
HHC predicts  ${G_E(Q^2) / \sqrt \tau G_M(Q^2)} \to 0 $ at large
$Q^2.$  The new data from Jefferson Laboratory~\cite{Jones:uu}
which shows a decrease in the ratio ${G_E(Q^2)/  G_M(Q^2)} $ is
not itself in disagreement with the HHC prediction.

The leading-twist QCD motivated form $Q^4 G_M(Q^2) \simeq {{\rm
const} / Q^4 \ln Q^2\Lambda^2}$ provides a good guide to both the
time-like and spacelike proton form factor data at $Q^2
> 5$ GeV$^2$ \cite{Ambrogiani:1999bh}.   However, the Jefferson
Laboratory data~\cite{Jones:uu} appears to suggest $Q
F_2(Q^2)/F_1(Q^2) \simeq {\rm const},$  for the ratio of the
proton's  Pauli and Dirac form factors in contrast to the nominal
expectation $Q^2 F_2(Q^2)/F_1(Q^2) \simeq {\rm const}$ expected
(modulo logarithms) from PQCD.  It should however be emphasized
that a PQCD-motivated fit is not precluded.  For example,  the form
\begin{equation}
{F_2(Q^2)\over F_1(Q^2)} = {\mu_A \over 1 + (Q^2/c) \ln^b(1+
Q^2/a)}
\end{equation}
 with
$\mu_A = 1.79,$ $a = 4 m^2_\pi = 0.073~$GeV$^2,$ $ b = -0.5922,$ $
c = 0.9599~$GeV$^2$ also fits the data well~\cite{BHHK}.

It is usually assumed that a heavy quarkonium state such as the
$J/\psi$ always decays to light hadrons via the annihilation of
its heavy quark constituents to gluons.  However, as Karliner and
I \cite{Brodsky:1997fj} have shown, the transition $J/\psi \to
\rho \pi$ can also occur by the rearrangement of the $c \bar c$
from the $J/\psi$ into the $\ket{ q \bar q c \bar c}$ intrinsic
charm Fock state of the $\rho$ or $\pi$.  On the other hand, the
overlap rearrangement integral in the decay $\psi^\prime \to \rho
\pi$ will be suppressed since the intrinsic charm Fock state
radial wavefunction of the light hadrons will evidently not have
nodes in its radial wavefunction.  This observation provides a
natural explanation of the long-standing puzzle why the $J/\psi$
decays prominently to two-body pseudoscalar-vector final states in
conflict with HHC, whereas the $\psi^\prime$ does not.  If the
intrinsic charm explanation is correct, then this mechanism will
complicate the analysis of virtually all heavy hadron decays such
as $J/\psi \to p \bar p.$ In addition, the existence of intrinsic
charm Fock states, even at a few percent level, provides new,
competitive decay mechanisms for $B$ decays which are nominally
CKM-suppressed~\cite{Brodsky:2001yt}.  For example, the weak
decays of the B-meson to two-body exclusive states consisting of
strange plus light hadrons, such as $B\to\pi K,$ are expected to
be dominated by penguin contributions since the tree-level $b\to s
u\bar u$ decay is CKM suppressed.  However, higher Fock states in
the B wave function containing charm quark pairs can mediate the
decay via a CKM-favored $b\to s c\bar c$ tree-level transition.
The presence of intrinsic charm in the $b$ meson can be checked by
the observation of final states containing three charmed quarks,
such as $B \to J/\psi D \pi$ \cite{Chang:2001yf}.

\section{TIMELIKE VIRTUAL COMPTON SCATTERING}

\begin{figure}[htb]
\begin{center}
\includegraphics*[width=75mm]{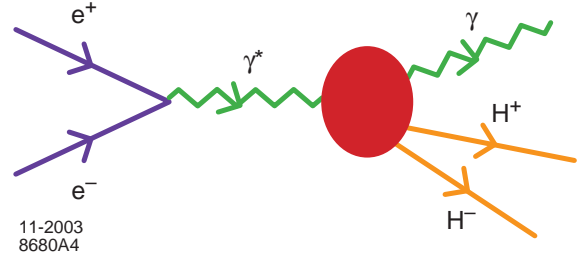}
\end{center}
\caption[*]{Process for measuring the timelike Compton  amplitude
$\gamma^* \to H \bar H \gamma$.} \label{sar8680A4}
\end{figure}

The Compton amplitude $\gamma \pi \to \gamma \pi$  is the simplest
two-body scattering amplitude in QCD after lepton-meson
scattering.  Despite its fundamental importance, the meson Compton
amplitude has never been measured directly.  However, one can make
interesting measurements of the timelike Compton amplitude using
$e^+ e^- \to \gamma^* \to H \bar H \gamma.$ See Fig.
\ref{sar8680A4}.  More generally, one can use electron-positron
annihilation to measure the timelike Compton amplitude for
virtually any hadron: $\gamma^* \to H \bar H \gamma$ where $H$ can
be any neutral or charged meson or baryon.  The interference with
the radiative return amplitude $e^+ e^- \to e^+ e^-  \gamma \to
\gamma^* \gamma \to H \bar H \gamma,$ which is proportional to the
timelike form factor $F_H(q^2),$ can be measured through $H
\leftrightarrow \bar H$ charge and single-spin asymmetries.  One
can also measure timelike transition Compton amplitudes $\gamma^*
\to H \bar H^*$ and timelike form factors.  In principle, the
spacelike and timelike amplitudes are related by crossing and
dispersion theory to generalized parton distributions; in
practice, the timelike analysis involves even more complexities
than virtual Compton scattering.  One of the most interesting
measures is the two-hadron distribution amplitude $\phi_{H \bar
H}(x,{\cal M}^2, \widetilde Q^2)$ which measures the transition
between a $q \bar q$ state and the $H \bar H$ hadron pair with
invariant mass ${\cal M}^2 = (p_H + p_{\bar
H})^2$~\cite{Muller:1994fv,Diehl:2000uv}. One can factorize this
distribution amplitude from the timelike virtual Compton amplitude
when the quark propagator has high virtuality $\widetilde Q^2.$ It
obeys the same operator product expansion and the same type of
logarithmic $\widetilde Q^2$ evolution as the pion distribution
amplitude.

The $\gamma^* \gamma \to \pi^+ \pi^-$ hadron pair process is
related to virtual Compton scattering on a pion target by
crossing.  The leading-twist amplitude is also sensitive to the $1/x -
1/(1-x)$ moment of the two-pion distribution amplitude coupled to
two valence quarks.

The virtual Compton scattering amplitudes $T(\gamma^*
\to H  \bar H \gamma)$ have extraordinary sensitivity to
fundamental features of hadron
structure~\cite{Ji:inclusive,Radyushkin:1997ki,%
Guichon:1998xv,Vanderhaeghen:1998uc,%
Collins:1999be,Diehl:combined,%
Blumlein:2000cx,Penttinen:2000dg}.  Even though the final-state
photon is on-shell, the deeply virtual Compton process probes the
elementary quark structure of the hadron near the light cone as an
effective local current.  In the spacelike case, the scaling,
Regge behavior, and phase structure of deeply virtual Compton
scattering $\gamma^* p \to \gamma p$ have been discussed in the
context of the covariant parton model in Ref. \cite{BCG7273}.  The
interference of Compton and bremsstrahlung amplitudes gives an
electron-positron asymmetry in the ${e^\pm  p \to e^\pm \gamma p}$
cross section which is proportional to the real part of the
Compton amplitude~\cite{BCG7273}.

To leading order in $1/Q$, the deeply virtual Compton scattering
amplitude $\gamma^* p \to \gamma p$  factorizes as the convolution
in $x$ of the amplitude $t^{\mu \nu}$ for hard Compton scattering
on a quark line with the generalized Compton form factors
$H(x,t,\zeta),$ $ E(x,t,\zeta)$, $\widetilde H(x,t,\zeta),$ and
$\widetilde E(x,t,\zeta)$ of the target proton.  Here $x$ is the
light-cone momentum fraction of the struck quark, and $\zeta=
Q^2/2 P\cdot q$ plays the role of the Bjorken variable.  The form
factor $H(x,t,\zeta)$ describes the proton response when the
helicity of the proton is unchanged, and $E(x,t,\zeta)$ is for the
case when the proton helicity is flipped.  Two additional
functions $\widetilde H(x,t,\zeta),$ and $\widetilde E(x,t,\zeta)$
appear, corresponding to the dependence of the Compton amplitude
on quark helicity.  These ``skewed" parton distributions involve
non-zero momentum transfer, so that a probabilistic interpretation
is not possible.  However, there are remarkable sum rules
connecting the chiral-conserving and chiral-flip form factors
$H(x,t,\zeta)$ and $ E(x,t,\zeta)$ with the corresponding
spin-conserving and spin-flip electromagnetic form factors
$F_1(t)$ and $F_2(t)$ and gravitational form factors $A_{\rm
q}(t)$ and $B_{\rm q}(t)$ for each quark and anti-quark
constituent~\cite{Ji:inclusive}.  Thus deeply virtual Compton
scattering is related to the quark contribution to the form
factors of a proton scattering in a gravitational field.  All of
these form factors can be measured for timelike photons in
$\gamma^* \to H \bar H \gamma$ for protons as well as other
hadrons.

\section{CHARGE ASYMMETRIES IN TIMELIKE EXCLUSIVE REACTIONS}

\begin{figure}[htb]
\begin{center}
\includegraphics*[width=75mm]{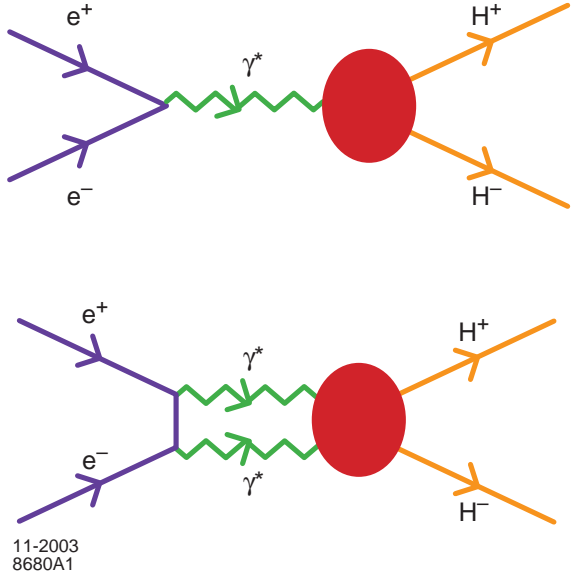}
\end{center}
\caption[*]{ Interference of one and two-photon exchange
amplitudes for $e^+ e^- \to H \bar H $ } \label{sar8680A01}
\end{figure}

The discrepancy between the Rosenbluth and polarization transfer
methods determinations of the proton form factors has led to a focus
on the role of two-photon exchange amplitudes
in $e p \to e p$ scattering.  In the timelike case, the interference
between the one- and two-photon exchange amplitudes in $e^+ e^-
\to H \bar H$ leads to a charge asymmetry at order $\alpha$
 a difference in the angular distribution of $H$ vs. $\bar H$ relative
to the incident electron direction.  See Fig. \ref{sar8680A01}.  This angular
asymmetry thus measures the relative phase of the $\gamma^* \gamma^* \to H \bar
H$ timelike Compton amplitude and the timelike form factor $\gamma^* \to H \bar
H.$ One also has to take into account the contribution to the asymmetry due to
the interference of amplitudes from soft photon radiation from the lepton and
hadron system.  One can use the charge asymmetry in $e^+ e^- \to \mu^+\mu^-$ as
the standard.

The theory of the two-photon exchange amplitude involves all of
the complexities of the doubly-virtual timelike Compton amplitude
$\gamma^* \gamma^* \to H \bar H.$  At high virtualities one
expects a quark handbag approximation~\cite{Kroll:2003cd} to be
valid.  The hadron asymmetry will then mimic the corresponding
$e^+ e^- \to \mu^+\mu^-$ asymmetry weighted by the sum of quark
charge squares and the $\langle{1\over x}\rangle$ $j=0$ moment
characteristic of a $j=0$ fixed pole in Regge
theory~\cite{Afanasev}.  A careful measurement of the charge
asymmetry in charged meson and baryon pair production could
illuminate the role of two-photon exchange in exclusive
amplitudes.

\section{THE PHOTON-TO-PION TRANSITION FORM FACTOR AND THE PION
DISTRIBUTION AMPLITUDE}

The simplest and perhaps most elegant illustration of an exclusive
reaction in QCD is the evaluation of the photon-to-pion transition
form factor $F_{\gamma \to \pi}(Q^2)$ which is measurable in
single-tagged two-photon $ee \to ee \pi^0$ reactions.  The form
factor is defined via the invariant amplitude $ \Gamma^\mu = -ie^2
F_{\pi \gamma}(Q^2) \epsilon^{\mu \nu \rho \sigma} p^\pi_\nu
\epsilon_\rho q_\sigma \ .$ As in inclusive reactions, one must
specify a factorization scheme which divides the integration
regions of the loop integrals into hard and soft momenta, compared
to the resolution scale $\widetilde Q$.  At leading twist, the
transition form factor then factorizes as a convolution of the
$\gamma^* \gamma \to q \bar q$ amplitude (where the quarks are
collinear with the final state pion) with the valence light-cone
wavefunction of the pion:
\begin{equation}
F_{\gamma M}(Q^2)= {4 \over \sqrt 3}\int^1_0 dx
\phi_M(x,\widetilde Q) T^H_{\gamma \to M}(x,Q^2) .
\label{transitionformfactor}
\end{equation}
The hard scattering amplitude for $\gamma\gamma^*\to q \bar q$ is
$ T^H_{\gamma M}(x,Q^2) = { [(1-x) Q^2]^{-1}}\left(1 + {\cal
O}(\alpha_s)\right). $ The leading QCD corrections have been
computed by Braaten~\cite{Braaten}.  The evaluation of the
next-to-leading corrections in the physical $\alpha_V$ scheme is
given in Ref. \cite{Brodsky:1998dh}.  For the asymptotic
distribution amplitude $\phi^{\rm asympt}_\pi (x) = \sqrt 3 f_\pi
x(1-x)$ one predicts $ Q^2 F_{\gamma \pi}(Q^2)= 2 f_\pi \left(1 -
{5\over3} {\alpha_V(Q^*)\over \pi}\right)$ where $Q^*= e^{-3/2} Q$
is the BLM scale for the pion form factor.  The PQCD predictions
have been tested in measurements of $e \gamma \to e \pi^0$ by the
CLEO collaboration~\cite{Gronberg:1998fj}.  The flat scaling of
the $Q^2 F_{\gamma \pi}(Q^2)$ data from $Q^2 = 2$ to $Q^2 = 8$
GeV$^2$ provides an important confirmation of the applicability of
leading twist QCD to this process.  The magnitude of $Q^2
F_{\gamma \pi}(Q^2)$ is remarkably consistent with the predicted
form, assuming the asymptotic distribution amplitude and including
the LO QCD radiative correction with $\alpha_V(e^{-3/2} Q)/\pi
\simeq 0.12$.  One could allow for some broadening of the
distribution amplitude with a corresponding increase in the value
of $\alpha_V$ at small scales.  Radyushkin \cite{Radyushkin}, Ong
\cite{Ong} and Kroll \cite{Kroll} have also noted that the scaling
and normalization of the photon-to-pion transition form factor
tends to favor the asymptotic form for the pion distribution
amplitude and rules out broader distributions such as the
two-humped form suggested by QCD sum rules \cite{CZ}.

\begin{figure}[htb]
\begin{center}
\includegraphics*[width=75mm]{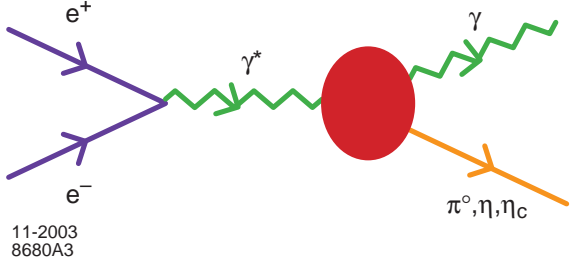}
\end{center}
\caption[*]{Process for measuring the timelike photon to meson transition
amplitude $\gamma^* \to M_0 \gamma.$} \label{sar8680A3}
\end{figure}

The photon-to-pion transition form factor $F_{\gamma \to
\pi}(q^2)$ is the simplest hadronic matrix element in QCD and also
one the most fundamental.  As noted above, the matrix element
$\VEV{\pi^0|j^\mu(0)|\gamma}$ transition form factor for spacelike
momenta has been measured in the spacelike domain $q^2 < 0$ by
scattering electrons on photons: $e \gamma \to e \pi^0.$ However,
$F_{\gamma \to \pi}(q^2)$ can also be measured in the timelike
domain  $q^2 =s > 0$ using $e^+ e^- \gamma^*\to \pi^0 \gamma.$ See
Fig. \ref{sar8680A3}.  Since the pion has positive $C$, there is
no background from radiative return.  Predictions for timelike
$q^2$ van be made by analytic continuation.  It would be very
valuable to test the PQCD predictions in the timelike domain,
including the effect of vector mesons in the approach to scaling.
One also can test predictions for the $\gamma \to H^0$ form factor
for any $C=+$ meson or hadronic system.  A comprehensive
discussion of the transition form factors for spacelike and
timelike $q^2$ is given in Ref.~\cite{Melikhov:2003hs}.

\section{EXCLUSIVE TWO-PHOTON ANNIHILATION INTO HADRON PAIRS}

\begin{figure}[htb]
\begin{center}
\includegraphics*[width=75mm]{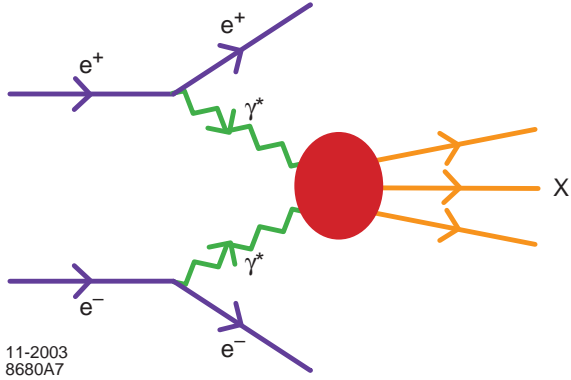}
\end{center}
\caption[*]{Illustration of two virtual photon annihilation in
lepton-lepton collisions.  The final state can be single $C=+$  hadrons,
hadron pairs (double virtual Compton), or more general $C=+$ systems. }
\label{sar8680A7}
\end{figure}

Two-photon reactions, $\gamma \gamma \to H \bar H$ at large s =
$(k_1 + k_2)^2$ and fixed $\theta_{\rm cm}$, provide a
particularly important laboratory for testing QCD since these
cross-channel Compton processes are the simplest calculable
large-angle exclusive hadronic scattering reactions involving two
hadrons.  See Fig. \ref{sar8680A7}.  The helicity structure, and
often even the absolute normalization can be computed for the
leading power-law contribution for each two-photon
channel~\cite{Brodsky:1981rp}.

In the case of meson pairs, dimensional counting predicts that for
large $s$, $s^4 d\sigma/dt(\gamma \gamma \to M \bar M$ scales at
fixed $t/s$ or $\theta_{\rm c.m.}$ up to factors of $\ln
s/\Lambda^2$.

The angular dependence of the $\gamma \gamma \to H \bar H$
amplitudes can be used to determine the shape of the
process-independent distribution amplitudes, $\phi_H(x,Q)$.  An
important feature of the $\gamma \gamma \to M \bar M$ amplitude
for meson pairs is that the contributions of Landshoff pitch
singularities are power-law suppressed at the Born level---even
before taking into account Sudakov form factor suppression.  There
are also no anomalous contributions from the $x \to 1$ endpoint
integration region.  Thus, as in the calculation of the meson form
factors, each fixed-angle helicity amplitude can be written to
leading order in $1/Q$ in the factorized form $[Q^2 = p_T^2 =
tu/s; \widetilde Q_x = \min(xQ,(l-x)Q)]$:
\begin{eqnarray}{\cal M}_{\gamma \gamma\to M \bar M}
&=& \int^1_0 dx \int^1_0 dy \phi_{\bar M}(y,\widetilde Q_y)\\
&&\times \ T_H(x,y,s,\theta_{\rm c.m.} \phi_{M}(x,\widetilde Q_x)
,\nonumber
\end{eqnarray}
where $T_H$ is the hard-scattering amplitude $\gamma \gamma \to (q
\bar q) (q \bar q)$ for the production of the valence quarks
collinear with each meson, and $\phi_M(x,\widetilde Q)$ is the
amplitude for finding the valence $q$ and $\bar q$ with light-cone
fractions of the meson's momentum, integrated over transverse
momenta $k_\perp < \widetilde Q.$ The contribution of non-valence
Fock states are power-law suppressed.  Furthermore, the
helicity-selection rules~\cite{Brodsky:1981kj} of perturbative QCD
predict that vector mesons are produced with opposite helicities
to leading order in $1/Q$ and all orders in $\alpha_s$.  The
dependence in $x$ and $y$ of several terms in $T_{\lambda,
\lambda'}$ is quite similar to that appearing in the meson's
electromagnetic form factor.  Thus much of the dependence on
$\phi_M(x,Q)$ can be eliminated by expressing it in terms of the
meson form factor.  In fact, the ratio of the $\gamma \gamma \to
\pi^+ \pi^-$ and $e^+ e^- \to \mu^+ \mu^-$ amplitudes at large $s$
and fixed $\theta_{CM}$ is nearly insensitive to the running
coupling and the shape of the pion distribution amplitude:
\begin{equation}
{{d\sigma \over dt }(\gamma \gamma \to \pi^+ \pi^-)
\over {d\sigma \over dt }(\gamma \gamma \to \mu^+ \mu^-)}
\sim {4 \vert F_\pi(s) \vert^2 \over 1 - \cos^2 \theta_{\rm c.m.} }
.\end{equation}

\begin{figure}[htbp]
\begin{center}
\includegraphics*[width=75mm]{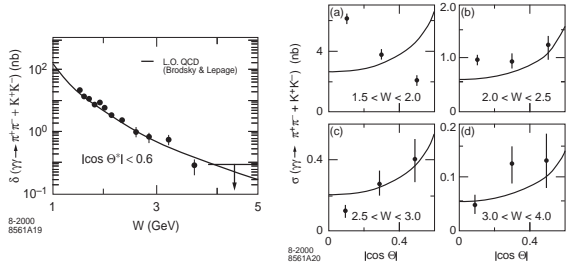}
\end{center}
\caption[*]{Comparison of the sum of $\gamma \gamma \rightarrow
\pi^+ \pi^-$ and $\gamma \gamma \rightarrow K^+ K^-$ meson pair
production cross sections with the scaling and angular
distribution of the perturbative QCD
prediction~\cite{Brodsky:1981rp}.  The data are from the CLEO
collaboration~\cite{Paar}.} \label{Fig:CLEO}
\end{figure}

The comparison of the PQCD prediction for the sum of $\pi^+ \pi^-$
plus $K^+ K^-$ channels with  CLEO data~\cite{Paar} is shown in
Fig.  \ref{Fig:CLEO}.  The CLEO data for charged pion and kaon
pairs show a clear transition to the scaling and angular
distribution predicted by PQCD~\cite{Brodsky:1981rp} for $W =
\sqrt{s_{\gamma \gamma}} > 2$ GeV.

It is  particularly important to measure the magnitude and angular
dependence of the two-photon production of neutral pions and
$\rho^+ \rho^-$ cross sections in view of the strong sensitivity
of these channels to the shape of meson distribution amplitudes.

Perturbative QCD predicts a strong suppression of the
leading-twist cross section for $\gamma \gamma \to \pi^0 \pi^0$
relative to the cross section for $\gamma\gamma \to \pi^+ \pi^-.$
This suppression is due to the negative interference between the
amplitudes involving two-quark currents with the single quark
current amplitudes.  This cancellation does not appear in models
based on the handbag approximation~\cite{Kroll:2003cd} in which
the only diagram which appears is a  factorized on-shell $\gamma
\gamma \to q \bar q$ Born amplitude. Thus the measurements of this
ratio is crucial for testing the perturbative QCD factorization of
exclusive amplitudes.  A similar test can be carried out by
measuring the neutral to charged pion pair ratio in $e^+ e^- \to
\pi \pi \gamma.$

QCD also predicts that the production cross section for charged
$\rho$-pairs (with any helicity) is much larger that for that of
neutral $\rho$ pairs, particularly at large $\theta_{\rm c.m.}$ angles.
Similar predictions are possible for other helicity-zero mesons.

Baryon pair production in two-photon annihilation is also an
important testing ground for QCD.  The only available data is the
cross channel reaction, $\gamma p \to \gamma p$.  The calculation
of $T_H$ for Compton scattering requires the evaluation of 368
helicity-conserving tree diagrams which contribute to $\gamma
(qqq) \to \gamma^\prime (qqq)^\prime$ at the Born level and a
careful integration over singular intermediate energy denominators
\cite{Farrar:1990qj,Kronfeld:1991kp,Guichon:1998xv}.  Brooks and
Dixon~\cite{Brooks:2000nb} have recently completed a recalculation
of the Compton process at leading order in PQCD, extending and
correcting earlier work.  It is useful to consider the ratio $
{s^6 d\sigma/dt(\gamma p \to \gamma p)/ t^4 F^2_1 (e p \to e p)}$
where $F_1(t)$ is the elastic helicity-conserving Dirac form
factor since the power-law fall-off, the normalization of the
valence wavefunctions, and much of the uncertainty from the scale
of the QCD coupling cancel.  The scaling and angular dependence of
this ratio is sensitive to the shape of the proton distribution
amplitudes and appears to be consistent with the distribution
amplitudes motivated by QCD sum rules.  The normalization of the
ratio at leading order is not predicted correctly by perturbative
QCD.  However, it is conceivable that the QCD loop corrections to
the hard scattering amplitude are significantly larger than those
of the elastic form factors in view of the much greater number of
Feynman diagrams contributing to the Compton amplitude relative to
the proton form factor.  The perturbative QCD predictions for the
phase of the Compton amplitude phase can be tested in virtual
Compton scattering by interference with Bethe-Heitler
processes~\cite{Brodsky:1972vv}.

Berger and Schweiger~\cite{Berger:2002vc} have recently  studied
baryon pair production in two-photon collisions using perturbative
QCD factorization  treating baryons as quark-diquark systems.
Their approach give a consistent description of the cross sections
for all octet baryon channels, including most recent
large-momentum-transfer data from LEP for the $\gamma \gamma \to
\Lambda \bar \Lambda.$ These prediction need to be compared with
the standard QCD analysis based on the three quark structure of
the baryons.

A debate has
continued~\cite{Isgur:1989iw,Radyushkin:1998rt,Bolz:1996sw,Diehl:2001dg}
on whether processes such as the pion and proton form factors and
elastic Compton scattering $\gamma p \to \gamma p$ might be
dominated by higher-twist mechanisms until very large momentum
transfer.  If one assumes that the light-cone wavefunction of the
pion has the form $\psi_{\rm soft}(x,k_\perp) = A \exp (-b
{k_\perp^2\over x(1-x)})$, then the Feynman endpoint contribution
to the overlap integral at small $k_\perp$ and $x \simeq 1$ will
dominate the form factor compared to the hard-scattering
contribution until very large $Q^2$.  However, this ansatz for
$\psi_{\rm soft}(x,k_\perp)$ has no suppression at $k_\perp =0$
for any $x$; {\em i.e.}, the wavefunction in the hadron rest frame
does not fall-off at all for $k_\perp = 0$ and $k_z \to - \infty$.
Thus such wavefunctions do not represent well soft QCD
contributions.  Endpoint contributions are also suppressed by the
QCD Sudakov form factor, reflecting the fact that a near-on-shell
quark must radiate if it absorbs large momentum.  One can show
\cite{Lepage:1980fj} that the leading power dependence of the
two-particle light-cone Fock wavefunction in the endpoint region
is $1-x$, giving a meson structure function which falls as
$(1-x)^2$ and thus by duality a non-leading contribution to the
meson form factor $F(Q^2) \propto 1/Q^3$.  Thus the dominant
contribution to meson form factors comes from the hard-scattering
regime.

\section{THE DOUBLY-VIRTUAL TIMELIKE COMPTON AMPLITUDE}

One can measure the virtual Compton amplitude $T(\gamma^*_1
\gamma^*_2 \to H \bar H)$  as a function of  spacelike $q^2_1,
q^2_2$ and $s \ge 4 m^2_H$ in the two-photon reaction:
\begin{equation}
e^+ e^- \to e^+ e^- \gamma^*_1 \gamma^*_2 \to e^+ e^-  H \bar
H.\end{equation}
This should be a possible measurement at high
luminosity $e^+ e^-$ colliders, particularly for meson pairs.

Assuming that quark Compton scattering is dominant (and the $j=0$
mechanism is relevant), we can  predict the ratio of the leading
power-law contribution to the virtual Compton amplitude at large
$q^2_1$ and $q^2_2$  to the corresponding lepton pair production
amplitude \begin{eqnarray} R^{p \bar p}_{2 \gamma}(q^2_1, q^2_2,s)
&=& {T(\gamma^*_1 \gamma^*_2 \to H \bar H)\over
  T(\gamma^*_1 \gamma^*_2 \to \mu^- \mu^+)}\nonumber\\
  &=&  (e^2_u+e^2_d) \langle{1\over x_q}\rangle F_H(s)\ .
  \end{eqnarray}
The $C=+$ form factor $F_p(s)$ should be similar to the proton's timelike
Dirac form factor $F_1(s).$

Thus one can empirically check the theoretical assumptions
underlying the two-photon exchange amplitude which we need to
describe the radiative correction to elastic $ep $ scattering.  It
is also an important constraint on the timelike $s  \ge 4 M^2_H$
input to the  two-photon exchange amplitude which interferes with
the one-photon amplitude to give the charge asymmetry in $e^+ e^-
\to H \bar H.$

\section{PERTURBATIVE QCD CALCULATION OF BARYON FORM FACTORS}

The baryon form factor at large momentum transfer provides an
important example of the application of perturbative QCD to
exclusive processes.  Away from possible special points in the
$x_i$ integrations (which are suppressed by Sudakov form factors)
baryon form factors can be written to leading order in $1/Q^2$ as
a convolution of a connected hard-scattering amplitude $T_H$
convoluted with the  baryon distribution amplitudes.  The
$Q^2$-evolution of the baryon distribution amplitude can be
derived from the operator product expansion of three quark fields
or from the gluon exchange kernel.  Taking into account the
evolution of the baryon distribution amplitude, the nucleon
magnetic form factors at large $Q^2$, has the
form~\cite{Lepage:1980fj,Lepage:1979zb,Brodsky:1981kj}
\begin{eqnarray}
G_M(Q^2)&\rightarrow& {\alpha^2_s(Q^2)\over Q^4}\sum_{n,m} b_{nm}
\left({\rm log}{Q^2\over \Lambda^2}\right)^{\gamma^B_n+\gamma^B_n}
\nonumber \\[1ex] &\times&
\left[1+\mathcal{O}\left(\alpha_s(Q^2),{m^2\over
Q^2}\right)\right]\quad .
\end{eqnarray}
where the $\gamma^B_n$ are computable anomalous
dimensions~\cite{Peskin:1979mn} of the baryon three-quark wave
function at short distance, and the $b_{mn}$ are determined from
the value of the distribution amplitude $\phi_B(x,Q^2_0)$ at a
given point $Q_0^2$ and the normalization of $T_H$.
Asymptotically, the dominant term has the minimum anomalous
dimension.  The contribution from the endpoint regions of
integration, $x \sim 1$ and $y \sim 1,$ at finite $k_\perp$ is
Sudakov suppressed
\cite{Lepage:1979za,Lepage:1979zb,Lepage:1980fj}; however, the
endpoint region may play a significant role in phenomenology.

The proton form factor appears to scale at $Q^2 > 5 \ {\rm GeV}^2$
according to the PQCD predictions.  Nucleon form factors are
approximately described phenomenologically by the well-known
dipole form $ G_M(Q^2) \simeq {1 / (1+Q^2/0.71\,{~\rm GeV}^2)^2}$
which behaves asymptotically as $G_M(Q^2) \simeq (1 /Q^4)( 1- 1.42
{~\rm GeV}^2/ Q^2 + \cdots)\,.$ This suggests that the corrections
to leading twist in the proton form factor and similar exclusive
processes involving protons become important in the range $Q^2 <
1.4\ {\rm GeV}^2$.

Measurements for the timelike proton form factor using $\bar p p
\to e^+ e^-$ annihilation are reported in
Ref.~\cite{Andreotti:bt}.  The results are consistent with
perturbative QCD scaling.  The ratio of the timelike to spacelike
form factor depends in detail on the analytic continuation of the
QCD coupling, anomalous dimensions~\cite{Brodsky:1998dh}.

The shape of the distribution amplitude controls the normalization
of the leading-twist prediction for the proton form factor.  If
one assumes that the proton distribution amplitude has the
asymptotic form: $\phi_N = C x_1 x_2 x_3$, then the convolution
with the leading order form for $T_H$ gives zero!  If one takes a
non-relativistic form peaked at $x_i = 1/3$, the sign is negative,
requiring a crossing point zero in the form factor at some finite
$Q^2$.  The broad asymmetric distribution amplitudes advocated by
Chernyak and Zhitnitsky~\cite{Chernyak:1984bm,Chernyak:1989nv}
gives a more satisfactory result.   If one assumes a constant
value of $\alpha_s = 0.3$, and $f_N=5.3 \times 10^{-3}$GeV$^2$,
the leading order prediction is below the data by a factor of
$\approx 3.$ However, since the form factor is proportional to
$\alpha^2_s f^2_N$, one can obtain agreement with experiment by a
simple renormalization of the parameters.  For example, if one
uses the central value~\cite{Belyaev:sa} $f_N=8 \times 10^{-3}
$GeV$^2$, then good agreement is obtained~\cite{Stefanis:1999wy}.
The normalization of the proton's distribution amplitude is also
important for determining the proton's
lifetime~\cite{Berezinsky:qb,Brodsky:1983st}.

A useful technique for obtaining the solutions to the baryon
evolution equations is to construct completely antisymmetric
representations as a polynomial orthonormal basis for the
distribution amplitude of multi-quark bound states.  In this way
one obtain a distinctive classification of nucleon $(N)$ and Delta
$(\Delta)$ wave functions and the corresponding $Q^2$ dependence
which discriminates $N$ and $\Delta$ form factors.  More recently
Braun and collaborators have shown how one can use conformal
symmetry to classify the eigensolutions of the baryon distribution
amplitude~\cite{Braun:1999te}.  They identify a new `hidden'
quantum number which distinguishes components in the $\lambda=3/2$
distribution amplitudes with different scale dependence.  They are
able to find analytic solution of the evolution equation for
$\lambda=3/2$ and $\lambda=1/2$ baryons where the  two lowest
anomalous dimensions for the $\lambda=1/2$ operators (one for each
parity) are separated from the rest of the spectrum by a finite
`mass gap'.  These special states can be interpreted as baryons
with scalar diquarks.  Their results may support Carlson's
solution~\cite{Carlson:1986mm}  to the puzzle that the proton to
$\Delta$ form factor falls faster~\cite{Stoler:1993yk} than other
$p \to N^*$ amplitudes if the $\Delta$ distribution amplitude has
a symmetric $x_1 x_2 x_3$ form.

\section{SINGLE-SPIN POLARIZATION EFFECTS
AND THE DETERMINATION OF TIMELIKE PROTON FORM FACTORS}

Although the spacelike form factors of a stable hadron are real, the
timelike form factors have a phase structure reflecting the
final-state interactions of the outgoing hadrons.  In general, form
factors are analytic functions $F_i(q^2)$ with a discontinuity for
timelike momentum above the physical threshold $q^2> 4 M^2.$ The
analytic structure and phases of the form factors in the timelike
regime are thus connected by dispersion relations to the spacelike
regime~\cite{baldini,Geshkenbein74,seealso}.  The analytic form and
phases of the timelike amplitudes also reflects resonances in the
unphysical region $0 < q^2 < 4M^2$ below the physical
threshold~\cite{baldini} in the $J^{PC} = 1^{--}$ channel, including
gluonium states and di-baryon structures.

Any model which fits the spacelike form factor data with an analytic
function can be continued to the timelike region.  Spacelike form
factors are usually written in terms of $Q^2 = - q^2$.  The correct
relation for analytic  continuation can be obtained by examining
denominators in loop calculations in perturbation theory.  The
connection is $Q^2 \rightarrow q^2 e^{-i\pi}$, or
\begin{equation}
\ln Q^2 = \ln (-q^2) \rightarrow \ln q^2 - i\pi  \ .
\end{equation}
If the spacelike $F_2/F_1$ is fit by a rational function of
$Q^2$, then the form factors will be  relatively real in the timelike
region also.  However,  one in general gets a complex result from the
continuation.

At very large center-of-mass energies, perturbative QCD factorization
predicts diminished final interactions in $e^+ e^- \to H \bar H$, since
the hadrons are initially produced with small color dipole moments.
This principle of QCD color transparency~\cite{Brodsky:xz} is also an
essential feature~\cite{Bjorken:kk} of hard exclusive $B$
decays~\cite{Beneke:2001ev,Keum:2000wi}, and it needs to be tested
experimentally.

There have been a number of explanations and theoretically
motivated fits of the new Jefferson laboratory $F_2/F_1$ data.
Belitsky, Ji, and Yuan~\cite{belitsky02} have shown that factors
of $\log(Q^2)$ arise from a careful QCD analysis of the form
factors.  The perturbative QCD form $Q^2 F_2/ F_1 \sim \log^2
Q^2$, which has logarithmic factors multiplying the nominal
power-law behavior, fits the large-$Q^2$ spacelike data well.
Others~\cite{ralston,miller} claim to find mechanisms that modify
the traditionally expected power-law behavior with fractional
powers of $Q^2$, and they also give fits which are in accord with
the data.  Asymptotic behaviors of the ratio $F_2/F_1$ for general
light-front wave functions are investigated in~\cite{BHHK}.  Each
of the model forms predicts a specific fall-off and phase
structure of the form factors from $ s \leftrightarrow t$ crossing
to the timelike domain.  A fit with the dipole polynomial or
nominal dimensional counting rule behavior would predict no phases
in the timelike regime.

\section{TIMELIKE MEASURES}

The center-of-mass angular distribution provides the analog of the
Rosenbluth method for measuring the magnitudes of various helicity
amplitudes.  The differential cross section for $e^- e^+
\rightarrow B \bar B$ when $B$ is a spin-1/2 baryon is given in
the center-of-mass frame by
\begin{equation}                               \label{xsctn}
{d\sigma \over d\Omega} =
{\alpha^2 \beta  \over 4 q^2}
                            \ D \ ,
\end{equation}
where $\beta = \sqrt{1-4m_B^2/q^2}$ and $D$ is given by
\begin{equation} D =  |G_M|^2 \left(1+ \cos^2\theta \right) + {1\over \tau }\,
       |G_E|^2 \sin^2\theta  \ ;
\end{equation}
we have used the Sachs form factors~\cite{walecka}
\begin{eqnarray} G_M &=& F_1 + F_2 \ , \nonumber \\ G_E &=& F_1 + \tau F_2
\ , \end{eqnarray} with $\tau \equiv {q^2 / 4 m_B^2} > 1$.

As noted by Dubnickova, Dubnicka, and Rekalo, and by
Rock~\cite{d}, the existence of the $T-$odd single-spin asymmetry
normal to the scattering plane in baryon pair production $e^- e^+
\rightarrow B \bar B$ requires a nonzero phase difference between
the $G_E$ and $G_M$ form factors.  The phase of the ratio of form
factors $G_E/G_M$ of spin-1/2 baryons in the timelike region can
thus be determined from measurements of the polarization of one of
the produced baryons.  In a recent paper, Carlson, Hiller, and
Hwang and I have shown that measurements of the proton
polarization in $e^+  e^- \to p \bar p$ strongly discriminate
between the analytic forms of  models which have been suggested to
fit the proton $G_E/G_M$ data in the spacelike region.
Polarization observables can be used to completely pin down the
relative phases of the timelike form factors.  The complex phases
of the form factors in the timelike region make it possible for a
single outgoing baryon to be polarized in $e^- e^+ \rightarrow B
\bar B,$ even without polarization in the initial state.

There are three polarization observables, corresponding to
polarizations in three directions denoted $z$, $x$, and $y$,
respectively.  Longitudinal polarization ($z$) refers to the
polarization state parallel to the direction of the outgoing
baryon.  Sideways ($x$) means perpendicular to the direction of
the outgoing baryon but in the scattering plane.  Normal ($y$)
means normal to the scattering plane, in the direction of $\vec k
\times \vec p$ where $\vec k$ is the electron momentum and $\vec
p$ is the baryon momentum, with $x$, $y$, and $z$ forming a
right-handed coordinate system.

The polarization ${\cal P}_y$ does not require polarization in the
initial state and is~\cite{d}
\begin{equation}    \label{py} {\cal P}_y = {  \sin 2\theta
\, {\rm Im} G_E^*  G_M
      \over D \sqrt{\tau} }
                    =
      { (\tau - 1 )\sin 2\theta \, {\rm Im} F_2^* F_1
      \over D  \sqrt{\tau} }
                    \ .
\end{equation} The other two polarizations require initial state
polarization.  If the electron has polarization $P_e$
then~\cite{d}
\begin{equation}       \label{px} {\cal P}_x =
    -P_e {2 \sin\theta \, {\rm Re} G_E^* G_M \over D \sqrt{\tau} }
     \ ,
\end{equation}
and
\begin{equation}    \label{pz} {\cal P}_z =  P_e {2  \cos \theta
|G_M|^2 \over D}  \ .
\end{equation}
The sign of ${\cal P}_z$ can be determined from physical
principles.  Angular momentum conservation and helicity
conservation for the electron and positron determine that ${\cal
P}_z/P_e$ in the forward direction must be $+1$, verifying the
sign of the above formula.

The polarization measurement in $e^+ e^- \to  p \bar p$ will require a
polarimeter for the outgoing protons, perhaps based on a shell of a
material such as carbon which has a good analyzing power.  However,
timelike baryon-antibaryon production can occur for any pair that is
energetically allowed.  Baryons such as the $\Sigma$ and $\Lambda$ which
decay weakly are easier to study, since their polarization is
self-analyzing.

The polarization ${\cal P}_y$ is a manifestation of the T-odd observable
$\vec k \times \vec p \cdot \vec S_p$, with $\vec S_p$ the proton
polarization.  This observable is zero in the spacelike case, but need
not be zero in the timelike case because final state interactions can
give the form factors a relative phase.

One can also predict~\cite{BHtau} the single-spin asymmetry
${\cal P}_y$ for QED processes  such as $e^+ e^- \to \tau^+
\tau^-$ which is sensitive to the imaginary part of the timelike
Schwinger correction to the lepton anomalous moment and Pauli form
factor.

\begin{figure}[htb]
\begin{center}
\includegraphics[height=2.5in,width=75mm]{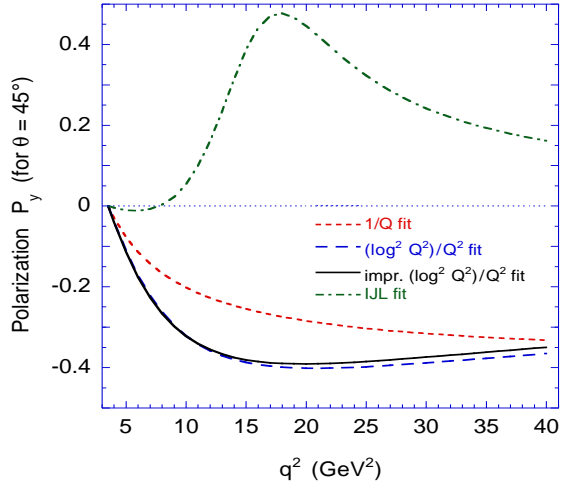}
\end{center}
\caption[*]{Predicted polarization ${\cal P}_y$ in the timelike
region for selected form factor fits described in the text.  The
plot is for $\theta = 45^\circ$.  The four curves are for an
$F_2/F_1 \propto 1/Q$ fit; the $(\log^2 Q^2)/Q^2$ fit of Belitsky
{\it et al.}; an improved $(\log^2 Q^2)/Q^2$ fit; and a fit from
Iachello {\it et al.}} Details are given in
Ref.~\cite{Brodsky:2003gs}. \label{figpy}
\end{figure}

Predictions  for polarization ${\cal P}_y$ in various models are
shown in Fig.~\ref{figpy}.  The predicted polarizations  are
significant and are distinct from a purely polynomial fit to the
spacelike data, which gives zero ${\cal P}_y$.

\begin{figure}[htb]
\begin{center}
\includegraphics[height=2.5in,width=75mm]{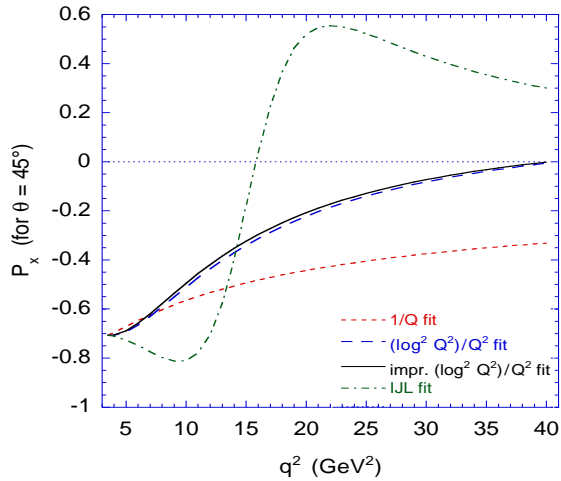}
\end{center}
\caption{The predicted polarization ${\cal P}_x$ in the timelike
region for $\theta=45^\circ$ and $P_e=1$.  The four curves correspond
to those in Fig.~\ref{figpy}.}
\label{fig:px}
\end{figure}

\begin{figure}[htb]
\begin{center}
\includegraphics[height=2.5in,width=75mm]{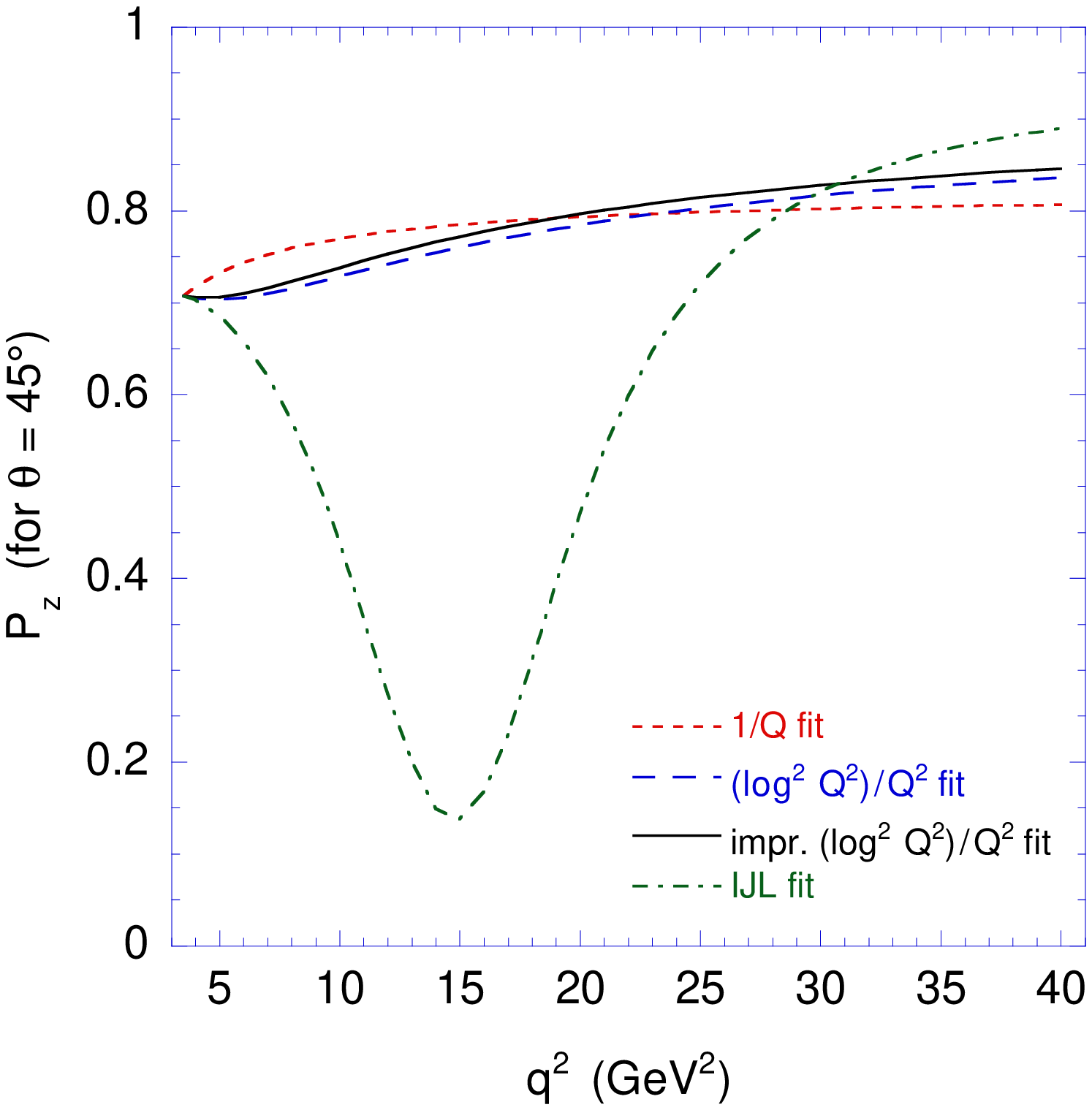}
\end{center}
\caption{The predicted polarization ${\cal P}_z$ in the timelike
region for $\theta=45^\circ$ and $P_e=1$.  The four curves
correspond to those in Fig.~\ref{figpy}. }
\label{fig:pz}
\end{figure}

The predictions for ${\cal P}_x$ and ${\cal P}_z$ are shown in
Figs.~\ref{fig:px} and~\ref{fig:pz}.  Both figures are for
scattering angle $45^\circ$ and $P_e = 1$.   The phase difference
$(\delta_E-\delta_M)$ between $G_E$ and $G_M$ is directly given by
the ${\cal P}_y / {\cal P}_x$ ratio,
\begin{equation} {{\cal P}_y \over {\cal P}_x} = {\cos\theta\over P_e}
    {{\rm Im\ } G_M^* G_E \over {\rm Re\ } G_M^* G_E}
    = {\cos\theta\over P_e} \tan(\delta_E-\delta_M)  \ .
\end{equation}

\begin{figure}[htb]
\begin{center}
\includegraphics[height=2.5in,width=75mm]{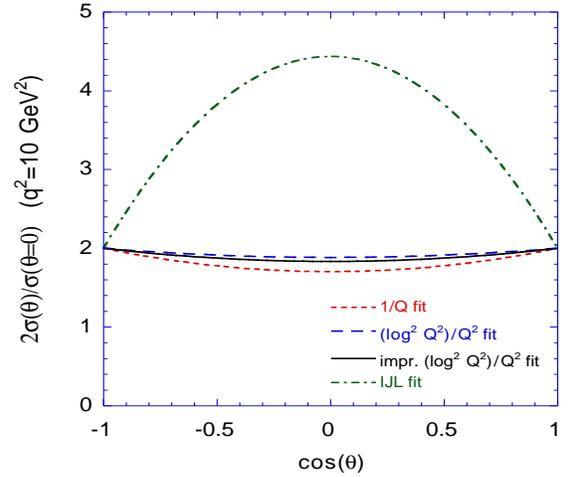}
\end{center}
\caption{The predicted differential cross section
$\sigma(\theta)\equiv d\sigma/d\Omega$.  The four curves correspond
to those in Fig.~\ref{figpy}.} \label{fig:diff}
\end{figure}

The magnetic form factor in the IJL model~\cite{ijl} is very small
in the 10 to 20 GeV$^2$ region (taking the dipole form for
comparison) and has a zero in the complex plane near $q^2 = 15$
GeV$^2$.   This accounts for much of the different behavior of the
IJL model seen in the polarization plots.  The IJL ratio for
$G_E/G_M$ is large compared to the other three models, and this
strongly affects the angular behavior of the differential cross
section as shown in Fig.~\ref{fig:diff} for $q^2 = 10$ GeV$^2$.

\section{INCLUSIVE SINGLE-SPIN ASYMMETRIES}

Spin correlations provide a remarkably sensitive window to
hadronic structure and basic mechanisms in QCD.   Among the most
interesting polarization effects are single-spin azimuthal
asymmetries (SSAs) in semi-inclusive deep inelastic scattering,
representing the correlation of the spin of the proton target and
the virtual photon to hadron production plane: $\vec S_p \cdot
\vec q \times \vec p_H$~\cite{Avakian:2002td}.  Such asymmetries
are time-reversal odd, but they can arise in QCD through phase
differences in different spin amplitudes.

The most common explanation of the pion electroproduction
asymmetries in semi-inclusive deep inelastic scattering is that
they are related to the transversity distribution of the quarks in
the hadron $h_{1}$~\cite{Jaffe:1996zw,Boer:2001zw,Boer:2002xc}
convoluted with the transverse momentum dependent fragmentation
function $H^\perp_1$, the Collins function, which gives the
distribution for a transversely polarized quark to fragment into
an unpolarized hadron with non-zero transverse momentum
\cite{Collins93,Barone:2001sp,Ma:2002ns,Goldstein:2002vv,Gamberg:2003ey}.

The QCD final-state interactions (gluon exchange) between the
struck quark and the proton spectators in semi-inclusive deep
inelastic lepton scattering can produce Sivers-type single-spin
asymmetries which survive in the Bjorken
limit~\cite{Brodsky:2002cx,Collins,Ji:2002aa}.   The fragmentation
of the quark into hadrons is not necessary, and one has a
correlation with the production plane of the quark jet itself
$\vec S_p \cdot \vec q \times \vec p_q.$   The required matrix
element measures the spin-orbit correlation $\vec S \cdot \vec L$
within the target hadron's wave function, the same matrix element
which produces the anomalous magnetic moment of the proton, the
Pauli form factor, and the generalized parton distribution $E$
which is measured in deeply virtual Compton scattering.  Since the
same matrix element controls the Pauli form factor, the
contribution of each quark current to the SSA is proportional to
the contribution $\kappa_{q/p}$ of that quark to the proton
target's anomalous magnetic moment $\kappa_p = \sum_q e_q
\kappa_{q/p}$~\cite{Brodsky:2002cx}.  Avakian~\cite{Avakian:2002td}
has shown that the data from HERMES and Jefferson laboratory could
be accounted for by the above analysis.  However, more analyses
and measurements, especially azimuthal angular correlations, will
be needed to unambiguously separate the transversity and Sivers
effect mechanisms.

Physically, the final-state interaction phase arises as the
infrared-finite difference of QCD Coulomb phases for hadron wave
functions with differing orbital angular momentum.  The
final-state interaction effects can be identified with the gauge
link which is present in the gauge-invariant definition of parton
distributions~\cite{Collins}.  When the light-cone gauge is chosen,
a transverse gauge link is required.  Thus in any gauge the parton
amplitudes need to be augmented by an additional eikonal factor
incorporating the final-state interaction and its
phase~\cite{Ji:2002aa,Belitsky:2002sm}.  The net effect is that it
is possible to define transverse momentum dependent parton
distribution functions which contain the effect of the QCD
final-state interactions.    The same final-state interactions are
responsible for the diffractive component to deep inelastic
scattering, and that they play a critical role in nuclear
shadowing phenomena~\cite{Brodsky:2002ue}.

Measurements from Jefferson Lab~\cite{Avakian:2003pk} also show
significant beam single spin asymmetries in deep inelastic
scattering.  Afanasev and Carlson~\cite{Afanasev:2003ze} have
recently shown that this asymmetry is due to the interference of
longitudinal and transverse photoabsorption amplitudes which have
different phases induced by the final-state interaction between
the struck quark and the target spectators just as in the
calculations of Ref. \cite{Brodsky:2002cx}.  Their results are
consistent with the experimentally observed magnitude of this
effect.  Thus similar FSI mechanisms involving quark orbital
angular momentum appear to be responsible for both target and beam
single-spin asymmetries.

A related analysis shows that the initial-state interactions from
gluon exchange between the incoming quark and the target spectator
system will lead to leading-twist single-spin target spin
asymmetries in the Drell-Yan process $H_1 H_2 \to \ell^+ \ell^- X$
\cite{Collins:2002kn,BHS2}. Initial-state interactions also lead
to a $\cos 2 \phi$ planar correlation in unpolarized Drell-Yan
reactions \cite{Boer:2002ju}.

\begin{figure}[htb]
\begin{center}
\includegraphics*[width=75mm]{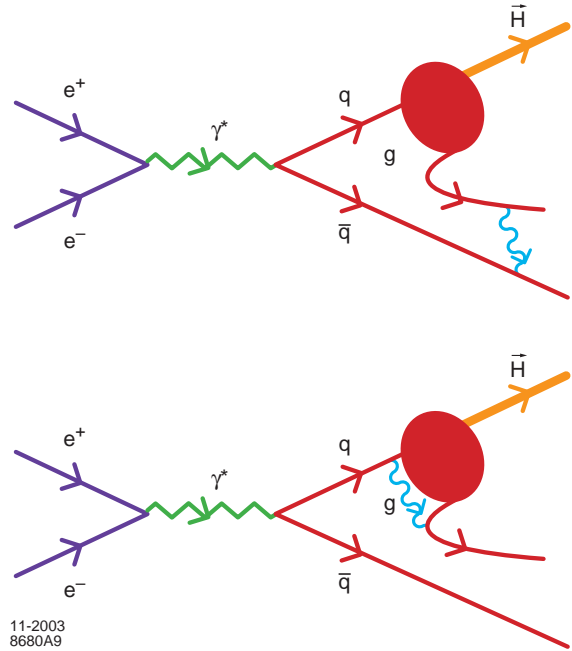}
\end{center}
\caption[*]{Illustration of the final-state gluon exchange
which produces single-spin asymmetries in inclusive electron-positron
collisions.}
\label{sar8680A09}
\end{figure}

We can also consider the SSA of $e^+e^-$ annihilation processes
for any inclusive process producing a  polarizable hadron, such as
$e^+e^-\to \gamma^* \to \pi {\Lambda} X$.  The $\Lambda$ reveals
its polarization via its decay $\Lambda \to p \pi^-$.  The final
state gluon exchange mechanism which causes $T-odd$ spin
correlations in inclusive $e^+ e^-$ annihilation processes is
illustrated in Fig. \ref{sar8680A09}.  The spin of the $\Lambda$
is normal to the decay plane.  Thus we can look for a SSA through
the T-odd correlation $\epsilon_{\mu \nu \rho \sigma}
S^\mu_\Lambda p^\nu_\Lambda q^\rho_{\gamma^*} p^\sigma_{\pi}$.
This is related by crossing to SIDIS on a $\Lambda$ target.  In
addition one can consider single spin asymmetries in inclusive
reactions such as $e^+e^-\to \gamma^* \to \pi X$ involving the
incident polarized electron beam.

\section{TESTING SOFT PION THEOREMS IN THE TIMELIKE DOMAIN}

In an important theoretical development, Pobylitsa {\em et
al.}~\cite{Pobylitsa:2001cz} have shown how to compute  transition
form factors linking the proton to nucleon-pion states which have
minimal invariant mass $W$.  A new soft pion theorem for high
momentum transfers allows one to compute the three-quark
distribution amplitudes for the near threshold pion states from a
chiral rotation.  The new soft pion results are in a good
agreement with the SLAC electroproduction data  for $W^2 <
1.4~$GeV$^2$ and $7 < Q^2 < 30.7~$GeV$^2.$

The soft pion analysis can be applied to  timelike reactions such as
$e^+ e^- \to p \bar n \pi^+$ in the regime where the pion is emitted at
small relative rapidity with respect to one of the outgoing nucleons.
The fall-off of the cross sections should be identical to that of $e^+ e^-
\to p \bar p.$

\section{NEAR-THRESHOLD COULOMB CORRECTIONS}

\begin{figure}[htb]
\begin{center}
\includegraphics*[width=75mm]{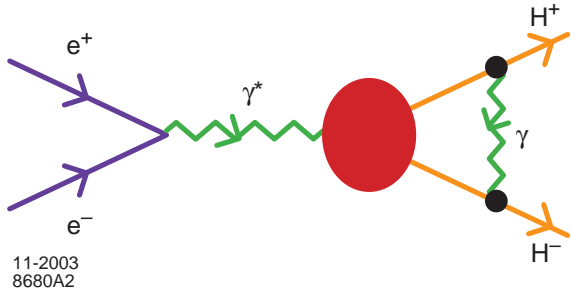}
\end{center}
\caption[*]{Final state Coulomb correction to charged hadron
pair production}
\label{sar8680A2}
\end{figure}

One of the most interesting effects due to QED radiative
corrections is the Coulomb correction to production of charges
pairs near threshold.  The lowest order Coulomb exchange is
illustrated in Fig. \ref{sar8680A2}.  The original theory is due
to Sommerfeld.  For example,
\begin{equation}
\sigma(e^+ e^- \to \bar p p)= \sigma_0(e^+ e^- \to \bar p p)
  {X\over 1-\exp{-X}}
\end{equation}
where $X= \pi \alpha \over \beta$, with $\beta^2= 1 - {4M^2_p\over
s}.$ Thus the absolute square of measured timelike form factors
$|G^p_M|^2$ and  $|G^p_E|^2$ are corrected by the factor ${X\over
1-\exp{-X}} \sim {\pi\alpha\over \beta}$ for small velocities
$\beta \ll  \pi \alpha.$  Thus the Coulomb correction becomes
infinite at zero relative velocity $\beta \to 0$!  The
Coulomb-corrected cross section is finite at threshold, although
the Born cross section vanishes linearly with $\beta$ due to the
vanishing phase space.  Observation of the angular distribution of
$\tau$ pair production can provide a  measurement of the magnetic
moment of the $\tau$~\cite{Brodsky:1995ds}.

The Coulomb enhancement in $e^+ e^- \to H^+ H^-$ is dual to the
presence of Coulomb $H^+ H^-$ bound states just below threshold.
In the case of $e^+ e^- \to \mu^+ \mu^-$ and $e^+ e^- \to \tau^+
\tau^-$ there is an accumulation of Bohr levels from ``true
muonium" $(\mu^+ \mu^-)$ and ``true tauonium" $(\tau^+ \tau^-)$
just below the continuum.

It would be interesting to observe these Coulomb
bound-state atoms.  In the case of $(p \bar p),$ $(\pi^+ \pi^-),$
$(D^+ D^-),$ {\em etc.}, the $S-wave$ Coulomb states decay
hadronically via annihilation, but the nonzero orbital states
could be quasi-stable.  It is possible that the threshold
enhancements seen in $p \bar p \to e^+ e^-$, and $J/\psi \to
\gamma p \bar p$ is due to the Coulomb enhancements.

\section{QCD THRESHOLD EFFECTS}

One can expect strong effects analogous to the QED Coulomb effects
whenever heavy quarks are produced at low relative velocity with
respect to each other or with  other quarks.  The opening of the
strangeness and charm threshold in timelike $e^+ e^-$ and $\gamma
\gamma$ reactions show sensitivity to this physics.  Two distinctly
different scales arise as arguments of the QCD coupling near
threshold: the relative momentum of the quarks governing the soft
gluon exchange responsible for the Coulomb potential, and a large
momentum scale approximately equal to twice the quark mass for the
corrections induced by transverse gluons.  One can use the angular
Distribution of heavy quarks to obtains a direct determination of
the heavy quark potential.  Predictions for the angular
distribution of massive quarks and leptons are presented in
Ref.~\cite{Brodsky:1995ds}, including the fermionic part of the
two-loop corrections to the electromagnetic form factors  using
with the BLM scale-fixing prescription.

\section{EFFECTIVE QCD CHARGES AND CONFORMAL ASPECTS OF QCD}

One can define the  coupling $\alpha_s$ of QCD from virtually any
physical observable~\cite{Grunberg:1980ja,Grunberg:1982fw}.  Such
couplings, called effective charges, are all-order resummations of
perturbation theory, so they  correspond to the complete theory of
QCD; it is thus guaranteed that they are analytic and
non-singular.  An important example is the effective charge
$\alpha_R$ where $1+ {\alpha_R(s)\over \pi}$ is defined from the
ratio of the total $e^+ e^-$ annihilation cross section to the
leading order QCD prediction.  Unlike the $\barMS$ coupling, a
physical coupling is analytic across quark flavor
thresholds~\cite{Brodsky:1998mf,Brodsky:1999fr}.  Furthermore, a
physical coupling must stay finite in the infrared when the
momentum scale goes to zero.  In turn, this means that integrals
over the running coupling are well defined for physical couplings.
Once such a physical coupling $\alpha_{\rm phys}(k^2)$ is chosen,
other physical quantities can be expressed as expansions in
$\alpha_{\rm phys}$ by eliminating the $\barMS$ coupling which now
becomes only an intermediary~\cite{Brodsky:1994eh}.  In such a
procedure there are in principle no further renormalization scale
($\mu$) or scheme ambiguities.  The physical couplings satisfy the
standard renormalization group equation for its logarithmic
derivative, ${{\rm d}\alpha_{\rm phys}/{\rm d}\ln k^2} =
\widehat{\beta}_{\rm phys}[\alpha_{\rm phys}(k^2)]$, where the
first two terms in the perturbative expansion of the Gell-Mann Low
function $\widehat{\beta}_{\rm phys}$ are scheme-independent at
leading twist, whereas the higher order terms have to be
calculated for each observable separately using perturbation
theory.

The  effective charge $\alpha_\tau(s)$ can be defined using the
high precision measurements of the hadronic decay channels of the
$\tau^- \to \nu_\tau {\rm h}^-$.  Let $R_{\tau}$ be the ratio of
the hadronic decay rate to the leptonic one.  Then $R_{\tau}\equiv
R_{\tau}^0\left[1+\frac{\alpha_\tau}{\pi}\right]$, where
$R_{\tau}^0$ is the zeroth order QCD prediction, defines the
effective charge $\alpha_\tau$.  The data for $\tau$ decays is
well-understood channel by channel, thus allowing the calculation
of the hadronic decay rate and the effective charge as a function
of the $\tau$ mass below the physical mass~\cite{Brodsky:2002nb}.
The vector and axial-vector decay modes which can be studied
separately.

Using an analysis of the $\tau$ data from the OPAL
collaboration~\cite{Ackerstaff:1998yj}, we have found that the
experimental value of the coupling $\alpha_{\tau}(s)=0.621 \pm
0.008$ at $s = m^2_\tau$ corresponds to a value of
$\alpha_{\MSbar}(M^2_Z) = (0.117$-$0.122) \pm 0.002$, where the
range corresponds to three different perturbative methods used in
analyzing the data.  This result is in good agreement with the
world average $\alpha_{\MSbar}(M^2_Z) = 0.117 \pm 0.002$.
However, from the figure we also see that the effective charge
only reaches $\alpha_{\tau}(s) \sim 0.9 \pm 0.1$ at $s=1\,{\rm
GeV}^2$, and it even stays within the same range down to
$s\sim0.5\,{\rm GeV}^2$.

The results for $\alpha_{\tau}(s)$ are  in good agreement with
the estimate of Mattingly and Stevenson~\cite{Mattingly:ej} for
the effective coupling $\alpha_R(s) \sim 0.85 $ for $\sqrt s <
0.3\,{\rm GeV}$ determined from ${\rm e}^+{\rm e}^-$ annihilation,
especially if one takes into account the perturbative commensurate
scale relation, $\alpha_{\tau}(m_{\tau^\prime}^2)= \alpha_R(s^*)$
where, for $\alpha_R=0.85$, we have $s^* \simeq
0.10\,m_{\tau^\prime}^2 .$ This behavior is not consistent with
the coupling having a Landau pole, but rather shows that the
physical coupling is close to constant at low scales, suggesting
that physical QCD couplings are effectively constant or ``frozen"
at low scales.  It is important to carefully extend the analysis of
$\alpha_R$ using annihilation data of higher precision and energy.

\begin{figure}[htb]
\begin{center}
\includegraphics*[width=75mm]{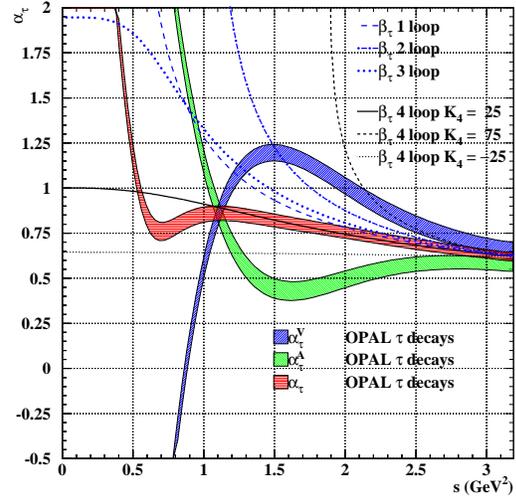}
\end{center}
\caption[*]{The effective charge
$\alpha_{\tau}$ for non-strange hadronic decays of a hypothetical
$\mathit{\tau}$ lepton with $\mathit{m_{\tau'}^2 = s}$ compared to
solutions of the fixed order evolution equation for
$\alpha_{\tau}$ at two-, three-, and four-loop order.  The error
bands include statistical and systematic errors.}
\label{fig:fopt_comp}
\end{figure}

Figure~\ref{fig:fopt_comp} compares the experimentally determined
effective charge $\alpha_{\tau}(s)$ with solutions to the
evolution equation for $\alpha_{\tau}$ at two-, {three-,} and
four-loop order normalized at $m_\tau$.  At three loops the
behavior of the perturbative solution drastically changes, and
instead of diverging, it freezes to a value $\alpha_{\tau}\simeq
2$ in the infrared.  The reason for this fundamental change is,
the negative sign of $\beta_{\tau,2}$.  This result is not
perturbatively stable since the evolution of the coupling is
governed by the highest order term.  This is illustrated by the
widely different results obtained for three different values of
the unknown four loop term $\beta_{\tau,3}$ which are also
shown\footnote{The values of $\beta_{\tau,3}$ used are obtained
from the estimate of the four loop term in the perturbative series
of $R_\tau$, $K_4^{\overline{\rm MS}} = 25\pm
50$~\cite{LeDiberder:1992fr}.} It is interesting to note that the
central four-loop solution is in good agreement with the data all
the way down to $s\simeq1\,{\rm GeV}^2$.

It has also been argued that $\alpha_R(s)$ freezes perturbatively
to all orders~\cite{Howe:2002rb}.  In fact since all observables
are related by commensurate scale relations, they all should have
an IR fixed point~\cite{Howe:2003mp}.  This result is also
consistent with Dyson-Schwinger equation studies of the physical
gluon propagator in Landau gauge~\cite{vonSmekal:1997is,Zwanziger:2003cf}.

The near constancy of the effective QCD coupling at small scales
helps explain the empirical success of dimensional counting rules
for the power law fall-off of form factors and fixed angle
scaling.  One can calculate the hard scattering amplitude $T_H$
for such processes~\cite{Lepage:1980fj} without scale ambiguity in
terms of the effective charge $\alpha_\tau$ or $\alpha_R$ using
commensurate scale relations~\cite{Brodsky:1998dh,Melic:2001wb}.
The effective coupling is evaluated in the regime where the
coupling is approximately constant, in contrast to the rapidly
varying behavior from powers of $\alpha_{\rm s}$ predicted by
perturbation theory (the universal two-loop coupling).  For
example, the nucleon form factors are proportional at leading
order to two powers of $\alpha_{\rm s}$ evaluated at low scales in
addition to two powers of $1/q^2$; The pion photoproduction
amplitude at fixed angles is proportional at leading order to
three powers of the QCD coupling.  The essential variation from
leading-twist counting-rule behavior then only arises from the
anomalous dimensions of the hadron distribution amplitudes.  The
magnitude of the  effective charge~\cite{Brodsky:1998dh}
$\alpha^{\rm exclusive}_s(Q^2) = {F_\pi(Q^2)/ 4\pi Q^2 F^2_{\gamma
\pi^0}(Q^2)}$ for exclusive amplitudes is connected to
$\alpha_\tau$ by a commensurate scale relation.  Its magnitude:
$\alpha^{\rm exclusive}_s(Q^2) \sim 0.8$ at small $Q^2,$  is
sufficiently large as to explain the observed magnitude of
exclusive amplitudes such as the pion form factor.

There are a number of useful phenomenological consequences of near
conformal behavior: the conformal approximation with zero $\beta$
function can be used as template for QCD
analyses~\cite{Brodsky:1985ve,Brodsky:1984xk} such as the form of
the expansion polynomials for distribution
amplitudes~\cite{Braun:1999te}. The near-conformal behavior of QCD
is also the basis for commensurate scale
relations~\cite{Brodsky:1994eh}  which relate observables to each
other without renormalization scale or scheme
ambiguities~\cite{Brodsky:2000cr}.  In this method the effective
charges of observables are related to each other in conformal
gauge theory; the effects of the nonzero QCD $\beta-$ function are
then taken into account using the BLM method~\cite{Brodsky:1982gc}
to set the scales of the respective couplings.  An important
example is the generalized Crewther relation~\cite{Brodsky:1995tb}
which allow one to calculate unambiguously without renormalization
scale or scheme ambiguity the effective charges of the polarized
Bjorken and the Gross-Llewellen Smith sum rules from the
experimental value for the effective charge associated with
$R_{e^+ e^-}(s).$ Present data are consistent with the generalized
Crewther relations within errors, but measurements at higher
precision  in $e^+ e^-$ annihilation are  needed to decisively
test these fundamental relations in QCD.  Such measurements are
also crucial for a high precision evaluation of the hadronic
corrections to the muon anomalous magnetic
moment~\cite{Davier:2003pw}.  The discrepancy between the
annihilation cross section in the isospin $I=1$ channel and the
corresponding isospin $I=1$ data from $\tau$ decay also needs to
be resolved~\cite{Ghozzi:2003yn}.

\section{Acknowledgements}

This talk was presented at the ICFA Workshop on Electron-Positron
Collisions in the $1$ to $2$ GeV Range, September 10-13, 2003 in
Alghero, Sardinia, Italy.  I thank Professor Rinaldo Baldini and
his colleagues at the Istituto Nazionale di Fisica Nucleare for
inviting me to this meeting.  I thank my collaborators,
particularly Carl Carlson, Guy de Teramond, Susan Gardner, Fred
Goldhaber, John Hiller, Dae Sung Hwang, Marek Karliner, Volodya
Karmanov, Jungil Lee,  Sven Menke, Carlos Merino, Johan Rathsman,
and Ivan Schmidt  for helpful discussions.  This work was
supported by the U.S. Department of Energy, contract
DE--AC03--76SF00515.

\end{document}